# On the impact of dataset size and class imbalance in evaluating machine-learning-based windows malware detection techniques

DAVID ILLES, Cambridge Spark, United Kingdom

**EXECUTIVE SUMMARY**

Cyber security is an increasingly relevant concern for organizations and governments across the globe. One of the contributing factors to this increased concern is the increasing difficulty to detect malware, especially the never seen before *zero-day* variety. This problem has led researchers to start applying machine learning techniques to malware detection and are reporting encouraging results.

The purpose of this project was to collect and analyse data about the comparability and real-life applicability of published results focusing on Microsoft Windows malware, more specifically the impact of dataset size and testing dataset imbalance on measured detector performance. These focus areas were identified because they are known challenges highlighted by prior academic enquiry. Some researchers use smaller datasets, and if dataset size has a significant impact on performance, that makes comparison of the published results difficult. Researchers also tend to use balanced datasets and accuracy as a metric for testing. The former is not a true representation of reality, where benign samples significantly outnumber malware, and the latter is approach is known to be problematic for imbalanced problems.

Apart from the importance of the problem to the academic discipline of cyber security due to investigating current challenges, the problem is important for organizations as well. Malware is one of the biggest threats facing businesses (Kleczynski, 2018) which disproportionately affects Microsoft Windows devices. Minor improvements in the identification of promising research directions can have a cascading effect on reducing cyber security losses and spend. Also, the problem has an environmental impact. Ransomware generates significant amount of bitcoin transactions because ransoms are most frequently paid in bitcoin (Paquet-Clouston *et al., 2019*), which is a technology that alone could push global warming above 2°C (Mora *et al.*, 2018).

The project identified two key objectives, to understand if dataset size correlates to measured detector performance to an extent that prevents meaningful comparison of published results, and to understand if good performance reported in published research can be expected to perform well in a real-world deployment scenario. These objectives were fulfilled through the utilisation of research questions, formulated as sensitivity analysis problems, investigating the relationship of detector performance to dataset size when detectors are trained and evaluated on balanced datasets using accuracy as a metric; dataset size when detectors are trained on a balanced dataset and evaluated simulating real-world usage; and testing set class imbalance when detectors are trained on a balanced dataset and evaluated simulating real-world usage.

The model of causality that informed this work was a simplistic model inspired by successionist and CMO approaches, that described the relationship of the research dataset to both real and estimated detector performance while also considering external factors.

The research was guided by the "positivism" research paradigm and even though it applied research questions instead of hypothesises, it heavily relied on a priori theory on machine learning and malware detection techniques to guide its scope. This was consistent with academic practice, as historically the vast majority of information security research has been quantitative using a positivist paradigm (Casaca and Florentino, 2014).

The research approach applied was primarily experimental, informed by a preliminary secondary research phase investigating popular methods, evaluation techniques, and results. This secondary research was conducted as a mini-literature review utilizing keyword-based search and Keshav's (2007) "three-pass approach". The main data collection experiment was conducted as a series of training/evaluation cycles of malware detectors. The detectors were trained on subsets of samples and features (format-agnostic, parsed, and their union) from the EMBER-2.0 dataset containing ~1 million samples (Anderson and Roth, 2018); sampling was conducted using a custom malware family-stratified historically accurate sampling; the classifier algorithms applied were Support Vector Machines, Random Forests, Decision Trees, and Boosted Algorithms; and the performance measures applied were accuracy and recall at a fixed (low) false positive rate. The experiment successfully generated accuracy scores including high 80%s and low 90%s on par with results from the literature, however Support Vector Machine-based detectors failed to produce reliable results due to computational issues.

All data generation, analysis, and visualisations were conducted using open-source python libraries providing the necessary flexibility to implement the required steps without constraints and ensured reproducibility through making the Python code available. To analyse the data visual techniques, Pearson's correlation coefficient, and repeated one-at-a-time local sensitivity analysis were used to deliver a simple to understand but well-rounded analysis producing visual, numerical, local, and global insights.

The research's results suggested that dataset size does correlate with measured detector performance to an extent that prevents meaningful comparison of published results, and without understanding the nature of the training set size-accuracy curve for published results (e.g.: at which point performance plateau) conclusions between approaches on which approach is "better" shouldn't be made solely based on accuracy scores. Results also suggested that high accuracy scores don't necessarily translate to high real-world performance; due to similar correlation between dataset-size and "real-world" performance, simply switching metrics from accuracy to recall at a fixed false positive rate, would not be sufficient on its own to facilitate a meaningful comparison; and that even if a more appropriate metric is used in combination with a high number of samples, the class imbalance ratio must be at least 1:8 to gain confidence that measured results will translate to similar performance during deployment. These findings have been consistent with and augment results from previous research conducted on other domains and Android malware detection.

As an outcome, further research directions were proposed for academia on techniques not covered by this project, and recommendations have been made both for academia and the cyber security practice more generally on how to interpret existing research and conduct further research efforts.

# TABLE OF CONTENTS

# GLOSSARY

**Accuracy**: a performance metric that represents the ratio of all correctly identified samples (both positives identified as positives and negatives identified as negatives) to all samples

**Area Under the Curve (AUC)**: a performance metric, the probability of a random positive sample ranked higher than a random negative sample (Tensorflow.org, 2021)

**Classifier**: a machine learning algorithm implementing discrete decision behaviours

**Dynamic analysis**: a type of software analysis which involves executing the program and monitoring its behaviour (Gibert et al, 2020)

**F-measure**: a performance metric, the harmonic mean of **Precision** and **Recall** (Wood, 2021)

**False Positive Rate (FPR)**: the percentage of false positives against all predicted positives

**Machine learning**: a type of technique that enables building systems that learn by example (Saxe and Sanders, 2018)

**Malware**: malicious software, a set of instructions that run on a computer and make the system do something that an attacker wants it to do (Skoudis and Zeltser, 2004)

**Precision**: a performance measure, the fraction of true positive examples among the examples that the model classified as positive (Wood, 2021)

**Ransomware**: a type of malware that prevents users from accessing certain resources until a monetary ransom is paid, (Mohanta and Saldanha, 2020)

**Recall**: a performance measure, the fraction of examples classified as positive among the total number of positive examples (Wood, 2021)

**Signature**: a small hash or byte-stream that contains information that identifies known-malware (Koret and Bachaakany, 2015)

**Static Analysis**: a type of software analysis which examines the code or structure of an executable without executing it (Gibert et al, 2020)

**True Positive Rate (TPR)**: same as recall

**Zero-day Malware**: a type of malware with no existing signatures to detect its presence

# 1 INTRODUCTION

Cyber security is an increasingly relevant concern for organizations and governments across the globe. A recent report from McAfee (in collaboration with the US Center for Strategic and International Studies) has found that in 2020 the cost of global cybercrime on global economy has exceeded $1 trillion (Smith and Lostri, 2020). This included high-profile incidents such as the ransomware attack on Colonial Pipeline, a US fuel pipeline system, which attack caused widespread disruptions including fuel shortages across the eastern seaboard (Gabbatt, 2021). A factor contributing to the success of attack campaigns is the increasing difficulty to detect malware such as ransomware. WatchGuard's Q2/2021 Threat Report (WatchGuard, 2021) reported that 91.5% of malware now arrives over encrypted connections, and 64.1% of all malwares falls into the category of *Zero-day Malware*, which means that no signatures exist to detect their presence. This trend has led researchers towards applying machine learning to malware detection with 7720 research papers published on the topic in 2018, a 98% increase from 2015 and a 476% increase with respect to 2010 (Gibert *et al*, 2020). Researchers are reporting great results using these techniques, for example (Kazi *et al*, 2019) has shown that decision trees can identify variations of the Zeus malware family with 93% precision. Considering this spike in interest and the potential impact of malware attacks, further enquiry into the interpretation of published results is timely and important.

The purpose of this project was to collect and analyse data about the comparability and applicability of published results, focusing on Microsoft Windows malware, and two key aspects of the problem domain:

- Impact of dataset size on measured detector performance
- Impact of testing dataset imbalance on measured detector performance

The project contributed to current theory by investigating known challenges.

A recent survey on the topic (Ucci *et al*, 2019) highlighted that dataset size is a recurring issue with published research. Many researchers use less than 1000 samples for evaluation, and only 39% of surveyed studies used more than 10,000 samples. If there is a significant impact on performance due to dataset size, that makes comparison of the published results difficult. This project provided insights to researchers on this problem that can guide the comparison process when results are surveyed.

Ucci has also highlighted that most surveyed papers benchmarked their results against balanced datasets, which scenario is unrealistic. As per Sophos (2021) around 3% of files are malware in a real setting, hence it is crucial to understand how detectors that perform well on balanced datasets handle more realistic distributions to better judge the real-world applicability of the proposed techniques. This project also provided insights about these applicability metrics and has the potential to help researchers consider published results in a wider context.

The project contributed to current practices by proposing recommendations on how to tackle both challenges which could be relevant not just for academia but industry practitioners building malware detection systems. This group might be wider than only anti-malware vendors, as the rise of "humanized" machine learning platforms increasingly empowers smaller and mid-sized businesses to harness advanced machine learning capabilities (Korda, 2019), enabling organisations to develop solutions for their unique variety of observed malware.

# 2 BACKGROUND

## 2.1 IDENTIFYING THE PROBLEM

The author identified the problem while investigating machine learning-based Android malware detection techniques during their studies for the Open University's M811 module (The Open University, 2022, M811). Having worked with fraud detection machine learning models in an industrial setting, the author found it interesting that Milosevic *et al.* (2017) used an extremely small sample of 400 apps and Wang *et al.* (2019) reported performance using accuracy scores, a problematic measure for imbalanced problems as it does not differentiate between error types. This motivated the author to investigate the problem further including other OSes like Windows, which investigation led them to literature including the Ucci *et al.* (2019) survey, confirming the topic to be a known challenge for the research area.

## 2.2 NATURE, EXTENT AND CHARACTERISTICS

Malware is a set of instructions that run on a computer and make the system do something that an attacker wants it to do (Skoudis and Zeltser, 2004). Since "Brain", the first self-replicating PC malware (a.k.a. "*virus*") appeared in 1986 (Leyden, 2006) malware has evolved to various types and functionalities that includes autonomously propagating *worms*, *keyloggers* that log the victim user's keystrokes, *banking malwares* that target financial information and credentials, *ransomware* that prevents users from accessing certain resources until a monetary ransom is paid, *cryptominers* that hijack the target's computing cycles to mine cryptocurrencies, and others (Mohanta and Saldanha, 2020).

As these behaviours are unwanted and harmful to the user, the detection of programs that exhibit such instructions became necessary, which lead to the creation of antivirus software. Early antivirus solely relied on the use of *signatures*, typically small hashes or byte-streams that contained information to identify known-malware (Koret and Bachaakany, 2015). Signatures had the advantage of producing small error rates, however the emergence of malware development toolkits like Zeus (Song *et al.*, 2008) enabled a massive proliferation of new malware armed with evasion techniques outpacing signature-based approaches, which led researchers to develop more robust solutions often based on machine learning (Ye *et al.*, 2017).

Machine learning techniques enable building systems that learn by example. Rather than building preconfigured rules, machine learning detection systems can be *trained* to determine whether a file is bad or good by learning from examples of good and bad files, automating the work of creating signatures with potential to perform more accurately than signature-based techniques, especially on previously unseen malware (Saxe and Sanders, 2018). The machine learning algorithms implementing such decision behaviours are also known as *classifier algorithms*. Testing the approach on data not included in the training examples is the last step of building these systems (Saxe and Sanders, 2018) which is crucial, as it generates performance metrics that suggests the real-world usability of the technique and provides a basis of comparison with competing approaches.

It is interesting to investigate certain characteristics of how performance is measured in recent literature focusing on *sample size*, *ratio of malware/benignware of samples,* and *performance metric used*. A recent survey of 40 papers on machine learning-based malware detection in executable files (Singh and Singh, 2021) has covered only 22 (55%) studies that leveraged more than 10,000 samples, only 5 (12.5%) that used *heavily imbalanced* (at least 10x more benignware than malware) datasets, with 36 (90%) using *Accuracy* as the sole reported performance metric. Another recent survey (Komatwar and Kokare, 2021) reporting on 34 papers did not disclose any details about the size/ratio of datasets used and reported only accuracy scores. Also, a survey of 65 papers (Ucci *et al.*, 2019) highlighted that only 39% of reviewed studies used more than 10,000 samples and that most surveyed papers benchmarked their results against balanced datasets, categorizing both as a known challenge in the field.

The sheer number of studies generating/reporting results this manner suggests that these known to be problematic practices affect the malware detection research area to a great extent.

Machine learning techniques are sensitive to dataset sizes. Evidence from other classification problems, for example Tweet Sentiment (Prusa *et al.*, 2015) suggests that significant performance improvements can be observed early in the dataset-size-to-performance curve before performance plateau and performance contribution of new samples diminishes. Knowing this inflection point for malware analysis techniques would be crucial to know when comparing results such as a 95.9% accuracy reported by (Ghiasi *et al.*, 2015) based on 1150 samples to a 99% accuracy from (Ali *et al.*, 2017) on 237,000 samples. Of course, dataset size is only one contributing factor to these results among: feature extraction techniques leveraged, qualitative characteristics of the dataset (e.g.: represented malware families), and others. This poses a challenge, however knowing the contributing factor of the dataset size would still provide value when performing similar comparisons.

*Accuracy* is a performance metric that represents the ratio of *all correctly identified samples* (both malwares identified as malware and benignware identified as benignware) to *all samples*, which has multiple problems. First, it doesn't differentiate between types of errors which is unrealistic, similarly to the medical industry where the ramifications of identifying cancerous patients as non-cancerous is overwhelmingly more costly than classifying a noncancerous patient as cancerous (Vluymans, 2019), the cost of falsely identifying malware as benign is not the same as the opposite. Second, ideally testing should be performed on sample distributions representing real use-cases. As per Sophos (2021) around 3% of the binary population is malware, in this setting accuracy is an unsuitable metric, as a trivially useless detector that never flags any binary as malware would immediately achieve a 97% accuracy score. Hence, results reporting accuracy scores would need to use a balanced testing dataset to have a chance of producing relevant results. As shown, most research follows this path, but evidence suggests that re-sampling the training set to achieve balance can lead to overfitting and information loss (Kaur *et al.*, 2019) so it is not trivial if good accuracy on a balanced dataset reliably translates to good performance in a real-world (highly imbalanced) setting. Evidence from research in the fraud detection domain also suggests that training class distribution affects the performance of the trained classifiers which can be mitigated by multi-classifier meta-learning approaches (Chan *et al.*, 1998), however most research surveyed use single-classifier approaches with unclear practical implications.

## 2.3 IMPORTANCE OF THE PROBLEM FOR ORGANIZATIONS

Malware is one of the biggest threats facing businesses (Kleczynski, 2018), disproportionately affecting Microsoft Windows devices. Windows has a 32.44% market share across all operating systems (Statcounter, 2021) targeted by 83.45% of all newly developed malware (Johnson, 2021) and 95% of identified ransomware (Virustotal, 2021). Microsoft spends over $1 billion annually on cyber security research and development (Cohen, 2017) to mitigate these threats, which cost is offset by the cost of not detecting malware effectively, exemplified by the estimated global $20 billion damage from ransomware in 2020 (PurpleSec, 2021). Due to these costs, minor improvements in the identification of promising research directions can have a cascading effect on reducing cyber security losses and spend, which should make the problem important to organizations.

## 2.4 IMPORTANCE OF THE PROBLEM SOCIALLY, ECONOMICALLY, ENVIRONMENTALLY

Malware is not only an organisational concern, better defences against zero-day malware could reduce the social and economic impact of attack campaigns. These at their extreme can be as significant as the devastation caused by the 2017 cyber-attacks on Ukraine, which affected the country's banks, power grid, postal service, government ministries, media organisations, the main airport in Kiev, nationwide mobile providers and even the Chernobyl power plant (Borys, 2017). Ransomware also generates significant amount of bitcoin transactions as ransoms are most frequently paid in bitcoin (Paquet-Clouston *et al., 2019*), a technology that alone could push global warming above 2°C (Mora *et al.*, 2018).

## 2.5 IMPORTANCE OF THE PROBLEM TO THE ACADEMIC DISCIPLINE OF CYBER SECURITY

As Ucci *et al.* (2019) highlighted, both proposed focus areas are known, current challenges. Considering the criticality of the topic, it is important for academics to understand the extent of these problems, however to the author's knowledge no previous research has investigated these exact problems specifically for Windows malware detection. Similar research has been conducted for Android Malware Detection by Zhao *et al.* ( 2021) who investigated the impact of sample duplication, and by Roy *et al.* (2015) and Allix *et al.* (2016) who explored the real-world applicability of published results. Unfortunately, none of these studies covered the sensitivity of results to dataset size, and even though Roy and Allix investigated the effects of real-world class imbalance on performance, Allix only leveraged a small malware dataset consisting of 1200 samples, not tested imbalance ratios larger than 1:3, and neither study contrasted their measurements with the frequently reported accuracy scores. This makes further academic enquiry into the topic desirable.

# 3 PROJECT EVALUATION AND SPECIFICATION

## 3.1 PERSONAL AND ACADEMIC SUITABILITY

The project was suitable for the researcher to take the role of an informed investigator. Apart from personal interest, the researcher has also: Worked with similar machine learning models in the payment fraud detection domain; Taught concepts related to model evaluation in a professional setting; Gained familiarity in the underlying cyber security concepts from the study of OU modules M811, M812, M817, T828; Had hands-on experience in reverse engineering from competing on CTF platforms like *HackTheBox.eu*; And had prior experience delivering projects of similar magnitude.

The project was suitable for other stakeholders. A brief stakeholder analysis that included considering motivations (Table-1), stakeholder mapping (Figure-1), and Mendelow power-interest analysis (Figure-2) (Mendelow, 1981) has shown that there are several interested stakeholders, some with high interest levels.

| Stakeholder | Motivation |
|---|---|
| Researcher | *Learn and pass the module* |
| Researcher's Family | *Safeguard the researcher's work-life balance* |
| Researcher's Employer | *Develop the researcher and increase knowledge* |
| Data Scientists | *Be aware of new results from their field* |
| Malware Analysts | *Understand/detect/stop malware* |
| Malware Developers | *Understand applied defences* |
| Cyber Security Leaders | *Protect cyber assets* |
| Anti-Malware Vendors | *Protect their client's assets* |
| T847 Tutor | *Provide guidance and assess performance* |
| T847 Students | *Learn and pass the module* |

Table-1: Stakeholders and their motivations

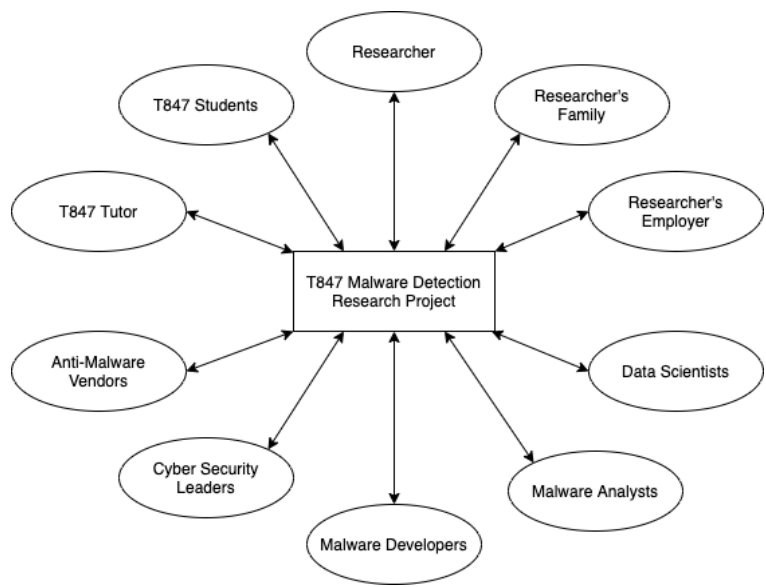

Figure-1: Stakeholder Map

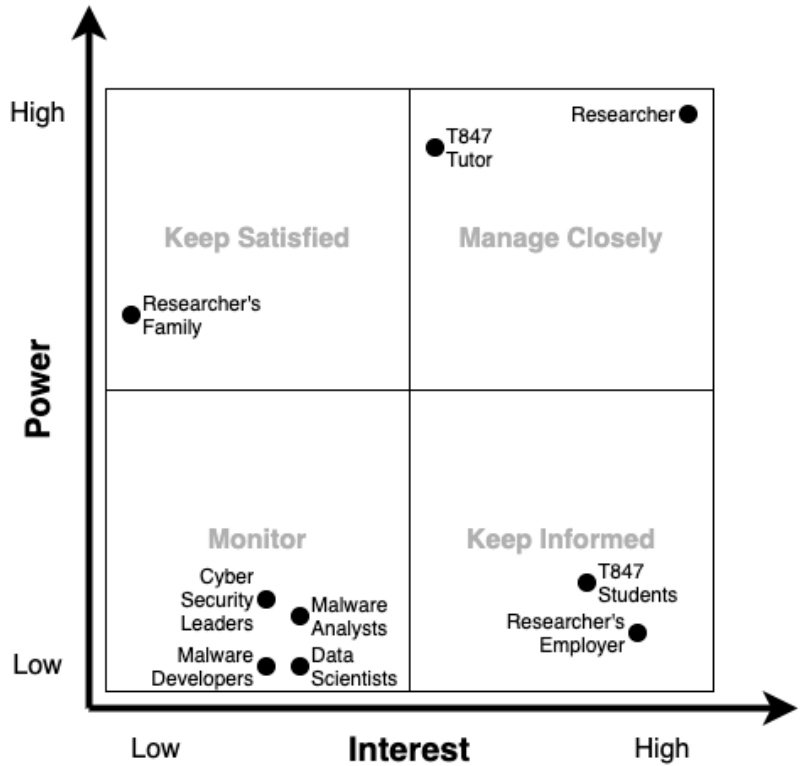
Figure-2: Mendelow Power-Interest Matrix

The research was suitable to the academic field of cyber security, as it was investigating known challenges outlined in 'Background' above.

## 3.2   AIM, OBJECTIVES, KEY TASKS & FEASIBILITY

The author conducted primary research to generate original data with the aim, objective and key tasks listed in Table-2 below:

| *Aim* | *To investigate how the results from published research on Windows machine learning-based malware detection techniques might be better analysed focusing on training dataset size and testing class imbalance regarding comparability and real-life applicability.* |
|---|---|
| *Objective #1* | *To understand if dataset size correlates to measured detector performance to an extent that prevents meaningful comparison of published results* |
| *Objective #2* | *To understand if good performance reported in published research can be expected to perform well in a real-world deployment scenario* |
| *Key Task #1* | *Survey literature to gain familiarity with popular methods, evaluation techniques and results* |
| *Key Task #2* | *Design data generation methodology informed by the surveyed literature* |
| *Key Task #3* | *Gather data on the impact of dataset size to detector performance as reported in literature* |
| *Key Task #4* | *Gather data on the impact of dataset size to estimated "real-world" performance* |
| *Key Task #5* | *Gather data on the impact of testing class imbalance to estimated "real-world" performance* |
| *Key Task #6* | *Analyse generated data* |
| *Key Task #7* | *Formulate recommendations* |

Table-2: Project aim, objectives, and key tasks

The outcomes were relevant for the researcher personally, the research stakeholders, and the academic discipline of cyber security for contextualising existing research results.

The scale and scope of the project was mainly determined by its data generation methods. It was possible to tailor this to the available time while retaining validity by finding a balance between the low-effort/low-fidelity extreme of tweaking existing models included with established benchmark datasets such and the high-effort/high-fidelity approach of manually collecting samples to be processed by custom processes.

The necessary non-OU resources for the project included: Time commitment from the researcher; Consultation time from workplace stakeholders; Software to conduct potential experiments and analysis; And computing power to perform these tasks.

## 3.3 RISK

A "Failure Mode and Effects Analysis" (FMEA) (The Open University, 2021, B1-A15) risk assessment was undertaken resulting in the following findings:

| Project Stage | Potential Failure | Potential Effect(s) | SR* | Potential cause of failure | LR* | PRN* | Prevention Plan | PEN* | PRF* |
| --- | --- | --- | --- | --- | --- | --- | --- | --- | --- |
| Detailed topic investigation | Missed significant relevant literature | Producing redundant or irrelevant results | 3 | Low effort or understanding | 3 | 9 | Study and advice solicitation | 0.3 | 2.7 |
|  | Disengaged stakeholders | Degraded quality or relevancy of results | 3 | Misunderstood office politics or poor interpersonal skills | 2 | 6 | Continuous stakeholder engagement | 0.2 | 1.2 |
| Research design | Unsuitable or unfeasible research plan | Repeated design stage | 6 | Insufficient skill | 5 | 30 | Study and advice solicitation | 0.3 | 9 |
| Research design documentation | Unsuitable academic style | Loss of marks | 4 | Low effort | 2 | 8 | Engagement with TMA feedbacks | 0.1 | 0.8 |
| Research preparation | Unavailability of required resources | Financial cost and/or extra work | 5 | Requirement miscalculation | 2 | 10 | Resourcing backup plans | 0.3 | 3 |
|  | Lack of input data | Inability to undertake research | 8 | Risky input data generation | 4 | 32 | Early focus on input data | 0.4 | 12.8 |
| Undertake research | Technical difficulties | Reduced available time | 3 | Unfamiliar tools | 2 | 6 | Preference of known tools | 0.1 | 0.6 |

| Analysis | Inconclusive results | Low research relevance | 6 | Insufficient skill | 5 | 30 | Study and advice solicitation | 0.3 | 9 |
|---|---|---|---|---|---|---|---|---|---|
| Findings and recommendation | Presented results benefiting malware developers more than analysts | Reputational loss, increase in cybercrime | 7 | Lack of ethics | 4 | 28 | Having a code of ethics | 0.2 | 5.6 |
| Writing-up | Missing hard EMA deadline | Failure | 10 | Unrealistic or non-existent schedule | 3 | 30 | Robust schedule | 0.1 | 3 |

*SR=Severity Ranking; LR=Likelihood Ranking; PRN=Priority Risk Factor; PEN=Plan Effectively Number; PRF=Residual Risk Factor*

Table-3: FEMA Analysis

## 3.4 PROJECT SPECIFICATION

### 3.4.1 PROJECT TITLE

On the Impact of Dataset Size and Class Imbalance in Evaluating Machine-Learning Based Windows Malware Detection Techniques

### 3.4.2 PROJECT SCHEDULE

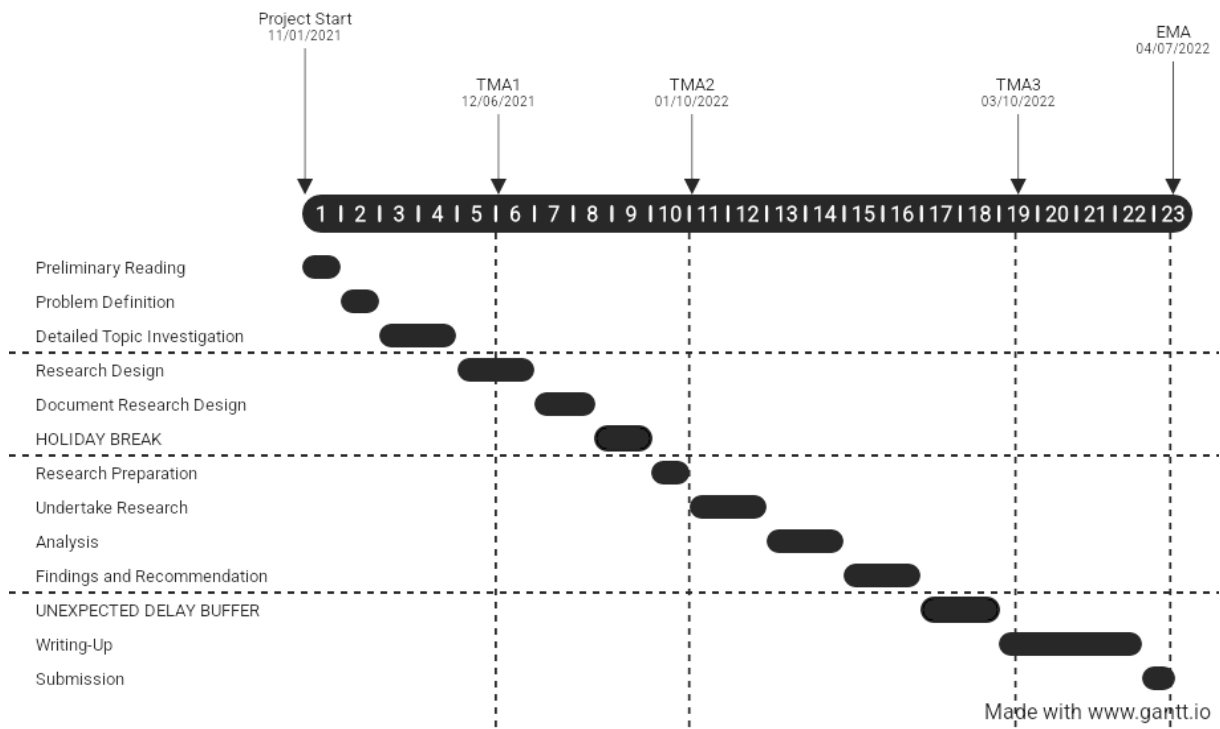

Figure-3: Project Schedule

# 4 THE RESEARCH PROCESS

## 4.1 RESEARCH QUESTIONS

To guide the research, research questions were formulated corresponding to data gathering Key Tasks #3, #4, and #5. This is consistent with the aim as answers to these questions synthetised with information obtained from executing Key Task #1 can directly provide answers to the research objectives, hence achieving the aim.

Research questions were preferred over hypothesises due their benefit of not requiring a priori outcome expectations that might bias the execution with the accepted drawback of potentially making analysis harder due to their wider focus.

The questions are:
1. What is the sensitivity of detector performance to dataset size when trained and evaluated on balanced datasets using accuracy as a metric?
2. What is the sensitivity of detector performance to dataset size when trained on a balanced dataset and evaluated simulating real-world usage?
3. What is the sensitivity of detector performance to testing set class imbalance when trained on a balanced dataset and evaluated simulating real-world usage?

The benefit of phrasing the questions as sensitivity analysis problems (instead of more general questions like "what is the impact of x on y?") was that the questions guided the research on how to deal with the inter-dependent nature of the outcome on many different inputs. This approach formalised the investigated phenomena into a model where hyperparameters (e.g.: Dataset, Features, etc…) are input to a process that ultimately produces performance as output (Figure-4) and enabled the application of established sensitivity analysis techniques. Since only one of the hyperparameters were investigated at a time, it had the drawback of potentially oversimplifying the problem if one or more inputs have complex relationships.

Restricting the approach to training on balanced datasets had the benefit of simulating the outcome of trying to productize published approaches as-is but had the drawback of excluding potential improvements from adjusting training practices.

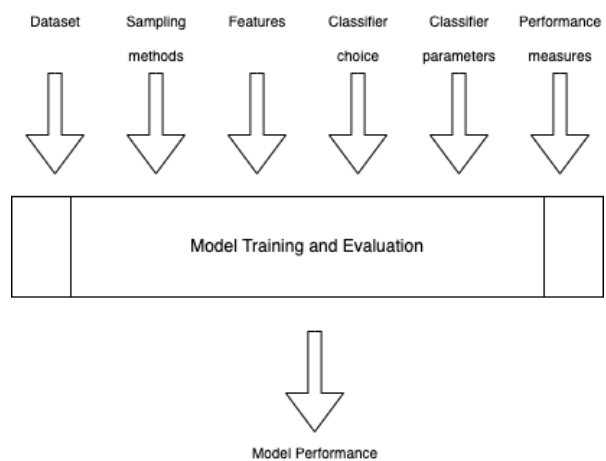

Figure-4: Model Training and Evaluation as a Model

## 4.2 MODEL OF CAUSALITY

The causal model that informed this work was a simplistic model based on Alves (2021) illustrated below on Figure-5:

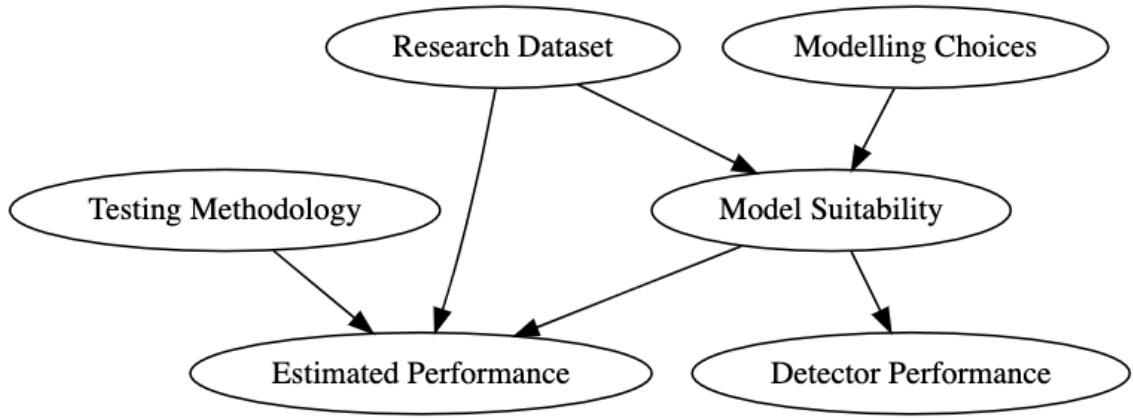

Figure-5: Model of Causality

The research assumed that it is plausible to assume that the cause-and-effect relationship between *Research Dataset* and *Estimated Performance* is significant enough to position it at the core of the research. As it is a single cause and effect relationship the research was arguably informed by a successionist causality model, however the research also accepted that other mechanisms play a role and need to be considered as context to ensure the validity of the data generation process, which arguably made the research a successionist-CMO hybrid.

There are multiple benefits to this model. Generally, it is a simplistic model. Specifically, it illustrates both the research's problem domain and some challenges associated with it well. It shows the causal relationship of the research dataset to detector performance and estimated performance, explaining why the research expected to find results when investigating dataset choices. It shows that there are external factors (Modelling Choices) that contribute to the outcomes the research needed to consider. But most importantly, it shows that there is no causal relationship from Estimated Performance to Detector Performance explaining the rationale of the research that questions the real-world performance of techniques that show good Estimated Performance. This was also an interesting challenge for the research itself, as the research aimed to simulate "real-world performance", but all the research could practically achieve is manipulate the Testing Methodology and ultimately could only measure Estimated Performance as well.

The key drawback of the model both generally and specifically is that it sacrificed some correctness for simplicity. Performance also depends on the environment (or "context) of the detector, which could have been captured by a realist/generative model (The Open University, 2021, B2-S2.3). Also, some of the causality is probabilistic due to the probabilistic behaviour of certain machine learning techniques (e.g.: there is a chance for unsuitable models to show high performance), which could have been captured by a probabilistic causal model (Hitchcock, 2018).

# 5 RESEARCH DESIGN AND METHODOLOGY

## 5.1 RESEARCH PARADIGM

The research was guided by the "positivism" research paradigm. Positivism presumes that knowledge is 'truth' that can be extracted from the empirical world through systematic, objective observation and measurement, which research if well planned and rigorously executed generates findings that can be generalized and replicated. Key features of positivism include the use of a priori theory or hypotheses to guide the scope of research, performing deductive research designed to test or refine prior knowledge, and carefully controlling conditions to prevent factors not part of the study affect findings. (The Open University, 2021, B2-S3.4)

This paradigm was congruent with the research's aim and objectives. Cyber space can be defined as a metaphysical construct created from the confluence of digital hardware (system), data, and humans (Edgar and Manz, 2017). This research's focus related to the intersection of system and data, both of which are well defined artificial constructs, hence truth should be independent of the observer, which the research aimed to extract through measurement with the aim to generalize for wider insights into comparability and applicability. Even though the research rejected the use of hypotheses in favour of research questions, it heavily relied on a priori theory on machine learning and malware detection techniques to guide its scope, and it was designed to test prior knowledge of detector performance while carefully controlling not studied factors. This was consistent with academic practice, as historically the vast majority of information security research has been quantitative using a positivist paradigm (Casaca and Florentino, 2014).

## 5.2 RESEARCH APPROACH/METHODOLOGY

The preferred approach for this research was experimental, collecting primary data through an experiment of repeated detector training and evaluation cycles. To inform creating a suitable experiment, a preliminary secondary research phase was also conducted to gain familiarity with popular methods, evaluation techniques and results.

Secondary research relies on reviewing and reanalysing existing (i.e., primary) data or information to produce new analytical or theoretical material (The Open University, 2021, B1-S7.2.2). In the context of this research the "product" of the secondary research phase was the experiment design used to generate the primary data.

Experiments are associated with the positivist paradigm and seek to investigate cause-and-effect relationships, which was the exact nature of the research questions. Elements of experimental design are an independent variable, its associated levels, control variables, a control procedure, and a dependent variable (Salkind, 2010). The independent variable (dataset size / testing set imbalance) is manipulated to test all associated levels while the control variables (features, parameters, etc.) are held constant to prevent them from impacting the dependent variable (detector performance), the subject of the data collection. The control procedure is the mechanism used to control the control variables, and in this computational experiment direct control over the variables was used to fix them at their desired values.

Another key feature of experiments is that they should be run more than once, where the repeat is expected to produce the same results and act as a check (The Open University, 2021, B2-S4). This reproducibility property for computational experiments can be defined as the ability of an independent group to obtain the same result using the author's artifacts (Ivie and Thain, 2018). The researcher's aim was to achieve this by making all source code available.

To make sure that the research did no harm to any human being, our society at large, or the academic discipline of cyber security, the research was guided by the ACM Code of Ethics and Professional Conduct (ACM, 2021).

## 5.3 RESEARCH APPROACH SUITABILITY

Machine learning is a field where most algorithms are too complex for formal analysis but provides experimental control over a wide range of factors (Langley, 1988). This property combined with an experiment's ability to provide quantitative results on a dependent variable while manipulating an independent variable is manipulated, made experimentation a suitable approach for this research. This choice was also in line with established practices, as all surveyed studies that form the basis of this research were also experimental.

Experiments can require a lot of time, effort, and expenditure (The Open University, 2021, B2-S4) making the approach in general unpractical for short projects. This concern was investigated in depth during the project's evaluation phase as part of the project's feasibility analysis, which concluded that the datasets, expertise, and computing power required to undertake this machine learning experiment were all available to the researcher, which made the approach not just suitable but also practical.

As an alternative approach, secondary research could have been conducted on the published results of surveyed papers, applying statistical methods to generate insights. This has the advantage of greatly simplifying data collection compared to conducting experiments and could have been a viable method to understand the correlation of dataset size to detector performance (Objective #1). Unfortunately, it would have introduced a significant risk of differences in other factors (e.g.: Modelling Choices) significantly skewing results, and it would have been necessary to augment it with another approach to understand real-world applicability (Objective #2) due to lack of published data. Since the most likely candidate for that augmenting approach would have been an experiment, it was considered more practical to cover both objectives with an experimental approach.

# 6 DATA GENERATION METHODS

## 6.1 DATA GENERATION TECHNIQUES

The secondary research outlined in Key Task #1 was conducted as a mini-literature review. A keyword-based search was conducted using the Open University Library (2022) and Google Scholar (Google, 2022), followed by filtering of results using the "three-pass approach" outlined by Keshav (2007). References of papers that reached the third pass were also reviewed.

- **Keywords:** malware, detection, classification, machine learning, analysis, survey

Building a detector as an experiment has a well-defined process: obtaining examples of malware/benignware samples (the dataset), extracting features from the examples, then training and testing the system (Saxe and Sanders, 2018). For each component of the process the below techniques were selected:

- **Dataset**: EMBER-2.0
- **Sampling Method**: Malware family-stratified historically accurate sampling
- **Classifier Algorithms**: Support Vector Machines, Random Forests, Decision Trees, Boosted Algorithms
- **Features:** Format-agnostic features, parsed features, and their union
- **Performance Measures:** Accuracy, Recall at 1% False Positive Rate

This selection is the outcome of Key Task #2 (informed by Key Task #1) and are used during Key Tasks #3, #4, #5. During these tasks a set of detectors for each research question were built and evaluated generating performance measures, which were analysed during Key Task #6.

## 6.2 JUSTIFICATION OF CHOICES

### 6.2.1 LITERATURE REVIEW

The scope of the literature review was a compromise between depth and the time available for conducting the project. The author considered including more sources to investigate and a deeper "snowballing" effect of reviewing references of papers identified through references, however it was deemed too risky that the inclusion of these techniques would not have left sufficient time for further research steps. The risk of missing significant results was mitigated by the fact that multiple recent surveys on the research topic were successfully identified as the part of the review.

### 6.2.2 DATASET

Lack of adequate large-scale public datasets is known challenge of the field (Akhtar, 2021). Obtaining malware can be done by manual collection or leveraging existing datasets. Manual collection can be performed either by using honeypots, or services like VirusTotal (2021). It provides the benefits of sample sizes only constrained by invested resources and access to fresh samples, which is important because detector performance can deteriorate as malware evolves (Galen and Seteele, 2020), however the technique could cost significant time and/or money. Existing datasets in contrast are readily available but might be outdated, limited in their size and/or quality, or might only contain pre-extracted features instead of full executables. Obtaining benignware is even more problematic due to copyright protections. Manual collection is still possible but require even more resources, however benchmark datasets almost exclusively contain pre-extracted features instead of full binaries. Even SOREL-20 that contains 10 million full (disarmed) malware executables (Harang and Rudd, 2020) contains only pre-extracted features for benign samples. Considering the timeline of this research and available resources, manual collection was deemed unfeasible, and the EMBER-2.0 dataset containing ~1 million samples (Anderson and Roth, 2018) was chosen as a compromise between relevancy and size.

### 6.2.3 SAMPLING METHODS

Since all research questions had specific requirements about the size (number of samples used) and structure (malware/benignware ratio) of the applied datasets, using the EMBER-2.0 dataset as-is was not possible, and some form of sampling was necessary. The simplest solution that could have satisfied these requirements was label-partitioned *stratified simple random sampling* where a sample is selected from each partition (stratum) (Anderson, 2021). The benefit of this technique is its simplicity however it also introduces problems. Evidence suggests that malware families can have a significant impact on detector performance (Wang *et al*, 2019) (different families might vary in identification difficulty, a mismatch in families present in training/testing datasets decreases detector performance) and ignoring malware timelines (e.g.: using malware knowledge "from the future" to detect in the present) can yield significantly biased results, both of which aspects the technique cannot control. Due to these problems the simple approach was rejected, and at the cost of higher complexity a custom stratified sampling solution was selected which also stratifies by malware families (achieving the same malware family distribution across training and testing datasets) and ensures that training samples historically precede testing samples.

### 6.2.4 CLASSIFIER ALGORITHMS

Regarding classifier algorithms, a balance needed to be found, as testing many classifiers have the benefit of providing more data, which strengthens the research's findings, however it increases the length and complexity of the research both during experimentation and data analysis. Figure-6 and Figure-7 show the results of a quick analysis on the distribution of classifier algorithms (excluding clustering) applied by papers surveyed in Singh and Singh (2021) Gibert *et al* (2020) and Ucci *et al* (2019):

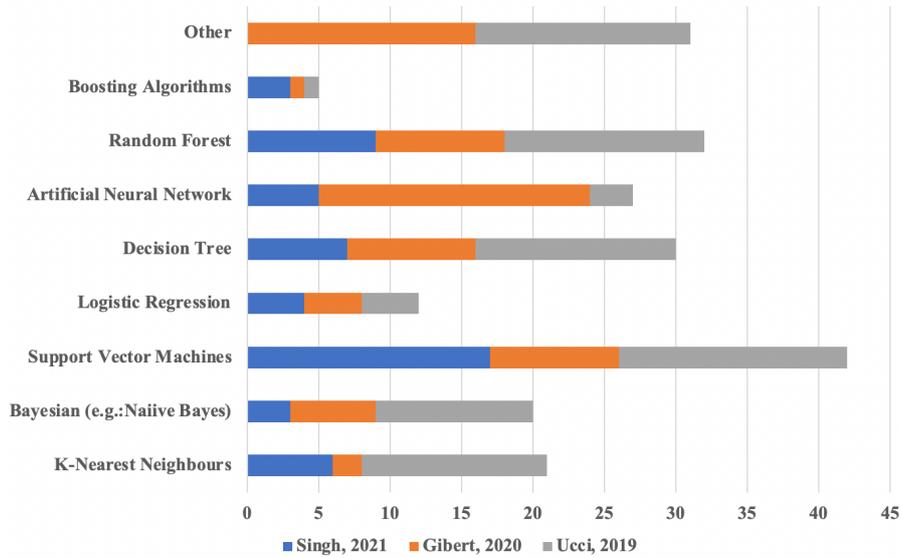

Figure-6: Surveyed Algorithms (Totals)

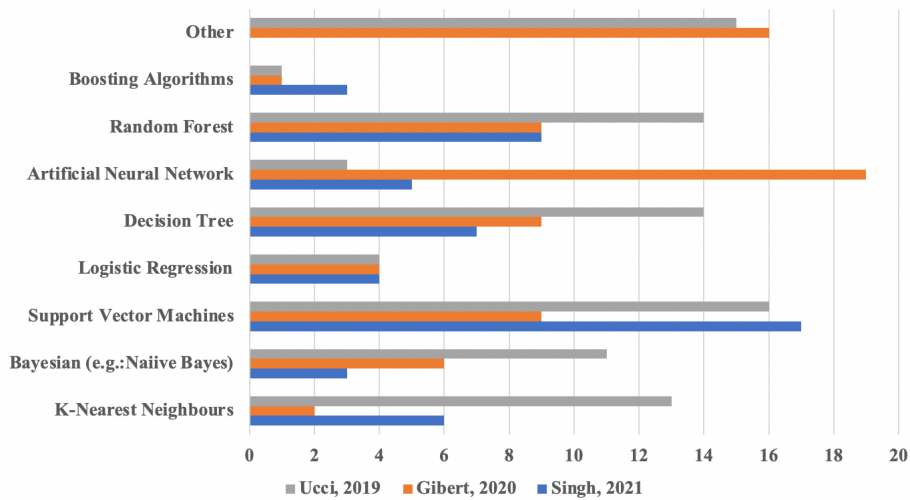

Figure-7: Surveyed Algorithms (Per Survey)

The results show that Support Vector Machines (SVM), Random Forest (RF), Decision Trees (DT) and Artificial Neural Networks (ANN) are amongst the most popular techniques applied in recent research. These were chosen to be evaluated, except for ANNs which were substituted with Boosting Algorithms (another robust technique) to reduce risk associated with the researcher's unfamiliarity with ANNs.

### 6.2.5 FEATURES

Malware detection features can be a result of performing static analysis, which examines the code or structure of an executable without executing it, or dynamic analysis, which involves executing the program and monitoring its behaviour (Gibert *et al*, 2020). Unfortunately, the EMBER dataset constrained the research, as it is limited to static analysis features (Anderson and Roth, 2018). Increasing the number of investigated feature sets with non-overlapping members increases the potential of generating well-generalizable insights, however each additional constructed model increases computing power requirements and the amount of data to be analysed. Also, the more partitioned the features are to different sets, the risk of not producing viable models also increases. To investigate more than a single approach, it was chosen to partition the EMBER features into two independently tested feature sets. Since the dataset already divides its features into two groups based on extraction method (parsed features vs. format-agnostic features) this division was accepted as a working model. To minimize the risk not achieving a viable model, the union of these feature sets were also selected.

### *6.2.6 PERFORMANCE MEASURES*

*Accuracy*, the *ratio of all correctly classified samples to all samples*, is a performance measure that needed to be applied as per the research question but is unsuitable when testing highly imbalanced distributions (Joshi, 2002), hence "real-world" evaluation needed a different measure. For imbalanced datasets the literature tends to recommend *F-measure* (Joshi, 2002), *the harmonic mean of Precision (the fraction of true positive examples among the examples that the model classified as positive) and Recall (the fraction of examples classified as positive among the total number of positive examples)* (Wood, 2021), or area under the "ROC" curve (*AUC*) (Jeni *et al*, 2013), *the probability of a random positive sample ranked higher than a random negative sample* (Tensorflow.org, 2021), however these can be problematic. Users find excessive false alerts unacceptable and may lead them to abandon the malware detector (Raff *et al*, 2020), hence in practice it should be maximized at an extremely low rate (Kaspersky.com, 2021). In this case AUC can be misleading as equal areas can greatly differ on individual false positive rates (FPR) (Figure-8), and F-measure is overcomplicated as one element of the harmonic mean is controlled. Considering the above, it was chosen to use *Recall at 1% FPR* as the performance measure for "real-world" evaluations. This had the benefits of providing a simple realistic measure but had the drawbacks of capturing results only at this arbitrarily set FPR.

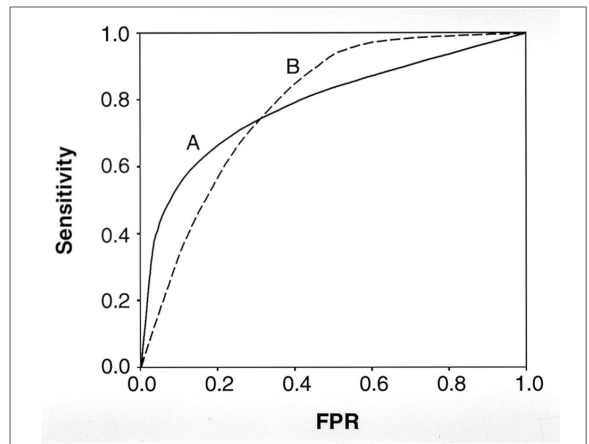

Figure-8: Crossing ROC curves with equal AUC (Park *et al*, 2004)

## 6.3 DATA GENERATION PLAN AND SCHEDULE

### 6.3.1 EXPERIMENTAL SETUP

Based on the above, 12 detectors were planned to be constructed (SVM, RF, DT, and Boosted models for parsed, format-agnostic and union feature sets) and trained/evaluated on datasets sampled from the EMBER-2.0 dataset using timeline-aware malware family-stratified sampling to generate data for the research questions. The planned training/evaluation details are summarized in Table-4 below:

| #Question | Training Set Size(s) [malware samples] | Training Set Class Ratio(s) [malware/benign] | Testing Set Size(s) [malware samples] | Testing Set Class Ratio(s) [malware/benign] | Performance Measure |
|---|---|---|---|---|---|
| #1 | 100, 1k, 5k, 10k, 50k, 100k | 1:1 | = Training Set | 1:1 | Accuracy |
| #2 | 100, 1k, 5k, 10k, 50k, 100k | 1:1 | 5k | 1:100 | Recall@1%FPR |
| #3 | 100k | 1:1 | 5k | 1:1, 1:5, 1:10, 1:25, 1:50 1:100 | Recall@1%FPR |

Table-4: Experimental Setup

### 6.3.2 COLLECTED DATA

The experiment was planned to generate a dataset of 216 observations (12 detectors * 3 experimental setups * 6 independent variable levels), each observation containing the performance measure from one run of the experiment. The observations were planned to be aggregated and analysed to extract insights about the relationship between the dependent and independent variables. The schema of the planned final dataset is described in Table-5 below:

| Column | Data Type |
|---|---|
| question | integer |
| algorithm | string |
| feature_set | string |
| train_set_size | integer |
| test_set_size | integer |
| test_set_ratio | float |
| perf_measure | string |
| performance | float |

Table-5: Observation dataset schema

### 6.3.3 DATA GENERATION SCHEDULE

Generating the data via the outlined experiment required a series of steps to be executed. Please find the original planned schedule of this execution (alongside the completion status as of the date the plan was first submitted for review as part of TMA2) on Figure-9 below:

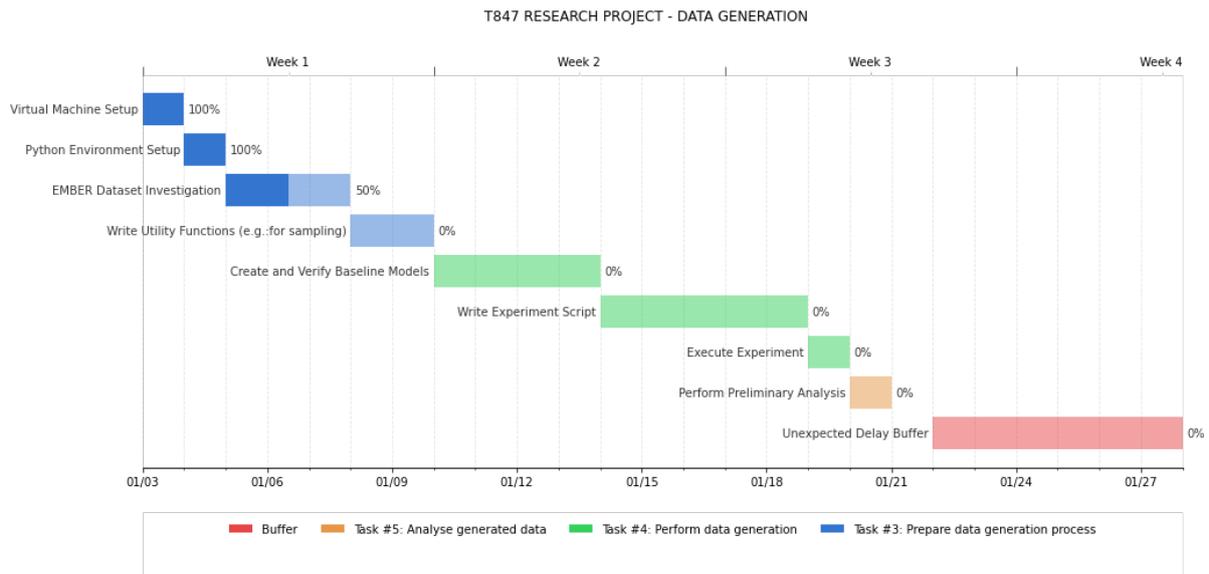

Figure-9: Data Generation Schedule

### 6.3.4 REPRODUCIBILITY

The experiment and subsequent analysis were performed using open-source data (Anderson and Roth, 2018) and open-source python packages available from PyPi.org (Python Software Foundation, 2022). This combined with making the experiment code available guarantees that any third party should be able to carry out the same investigation and reproduce the same results.

### 6.3.5 TRIANGULATION OF DATA

To improve the validity of the research the author primarily focused on "multiple methods of data collection" and "repeated observations over time" from the type of triangulation techniques highlighted by Bamberger *et al.* (2006). The former was implemented by the diversification of classifier algorithms and feature sets to be tested, and the latter was implemented in the execution stage as re-running the experiment with multiple random seeds once the computational requirements were known.

# 7 ASSESSMENT OF DATA GENERATION AND ANY RELATED ISSUES

The data generation closely followed the original plan of executing training/evaluations cycles of malware detectors on data sampled from the EMBER-2.0 dataset, successfully generating accuracy scores including high 80%s and low 90%s on par with results from literature, however some issues were encountered that resulted in minor revisions to the applied techniques:

1. The researcher was unable to successfully execute the feature vectorization script supplied with the dataset due to excessive memory consumption and chose to re-implement vectorization following the process described by Anderson and Roth (2018). Since the code is made available in the appendix, this does not hurt reproducibility, but will make it harder to relate other EMBER-2.0 based research due to potential implementation differences.

2. The dataset's malware family distribution couldn't satisfy historical precedence and strict malware family stratification between train/test sets at the same time for large samples. As a result, the stratification was relaxed to guaranteeing that both sets would contain samples from the same 50 malware families but at different ratios.

3. The dataset with the chosen 50 malware families couldn't supply enough malware for the most imbalanced (1:100) scenario. To get around the issue the number of malware samples in these datasets were reduced to 1500 samples.

4. During the experiment Decision Trees and Support Vector Machines did not produce Recall results at 1% False Positive Rate (FPR), the closest was ~10% FPR, which was accepted as practical reality for these detectors. This required the analysis stage to explicitly consider FPR value differences to ensure validity.

5. During preliminary analysis it was noted that the identified independent variables levels were unfortunate, as they didn't follow constant increments, and data point for some steps were not numerous enough for the preferred statistical techniques, posing threats to validity. To mitigate the issue, levels were refined (Table-6), and measurements were re-executed 3 times with different random seeds increasing the observation count from 216 to 1080. This increase should also strengthen the reliability of the results.

| #Q | Training Set Size(s) [malware samples] | Training Set Class Ratio(s) [malware/benign] | Testing Set Size(s) [malware samples] | Testing Set Class Ratio(s) [malware/benign] | Performance Measure |
|---|---|---|---|---|---|
| **#1** | 100, 200, 400, 800, 1.6k, 3.2k, 6.4k, 12.8k, 25.6k 51.2k, 102.4k | 1:1 | = Training Set | 1:1 | Accuracy |
| **#2** | 100, 200, 400, 800, 1.6k, 3.2k, 6.4k, 12.8k, 25.6k 51.2k, 102.4k | 1:1 | 1500 | 1:128 | Recall@1%FPR |
| **#3** | 102.4k | 1:1 | 1500 | 1:1, 1:2, 1:4, 1:8, 1:16, 1:32, 1:64, 1:128 | Recall@1%FPR |

Table-6: Revised Experimental Setup

Fortunately, none of the above issues had a serious impact on the data generation schedule. The original plan was robust enough to accommodate the extra necessary work for #1-#4, and #5 took 3 days from the unexpected delay buffer in the plan.

There were 2 issues identified that might impact the validity of the results:

1. The iteration count for SVM detectors had to be decreased for training to finish in reasonable times, however this caused the training to fail to converge, potentially resulting in sub-optimal performance.

2. Some accuracy scores generated were lower than scores reported in the literature, hence analysing sensitivity at these datapoints is arguably not an apples-to-apples comparison.

To overcome these issues the analysis mostly focused on results from non-SVM detectors and considered data points above 80% accuracy as more relevant.

# 8 ANALYSIS AND FINDINGS

## 8.1 TOOLS AND TECHNIQUES USED FOR DATA ANALYSIS

All data analysis were conducted using the NumPy (NumPy, 2021) and Pandas (NumFOCUS, 2022) statistical Python libraries, and all visualisations were created using the Seaborn (Waskom, 2021) Python library. These open-source tools provided the necessary flexibility to implement the required techniques without constraints and ensured reproducibility through making the Python code available. These benefits came at the cost of introducing the complexity of programming into the project, which considering the target audience was deemed acceptable.

The following techniques were used for data analysis:

- Visual techniques (Line Plots, Box Plots, and Regression Plots)
- Pearson's correlation coefficient
- Repeated one-at-a-time local sensitivity analysis

Visual techniques are effective ways to generate intuitions on trends and relationships in data. They have the benefits of being easy to produce and simple to understand by stakeholders who might be lacking deep statistical knowledge, however they lack the precision of numerical techniques. This technique is extensively used by similar research in the Android domain such as Allix *et al.* (2016).

Pearson's correlation is a technique to evaluate whether there is statistical evidence for a linear relationship among variables (Kent State University, 2022). It has the benefit of describing relationships with a single, comparable number. This complements the intuition given by visual techniques, but it can't describe localised differences in relationships at different variable levels.

To provide numerical results at different levels, the simple *one-way method* of sensitivity analysis was used at multiple independent variable levels. This method works by changing one factor (variable) by a fraction of its nominal value, holding all other factors (variables) constant, and observing the resultant fractional change in the output (Qian and Mahdi, 2020). An increment by factor of x2 was used for all data points, and the results were also further analysed with visual techniques. The technique's key benefit is its simplicity, but (similarly to Pearson's correlation) it is most suitable when the model is linear (Qian and Mahdi, 2020).

The combination of the above techniques can deliver a simple to understand but well-rounded analysis that produces visual, numerical, local, and global insights.

## 8.2 A NOTE ON PROJECT ARTIFACTS

The project has generated a significant number of observations and required an amount of python code which could not fit in the appendix of this report in their entirety. To work around this limitation, only a subset of the code and a small sample of the generated data was provided in the appendix, and the full dataset alongside the complete codebase was made available in a publicly accessible GitHub repository: *https://github.com/davidilles/msc-project-public*

## 8.3 ANALYSIS AND FINDINGS

### 8.3.1 QUESTION 1: WHAT IS THE SENSITIVITY OF DETECTOR PERFORMANCE TO DATASET SIZE WHEN TRAINED AND EVALUATED ON BALANCED DATASETS USING ACCURACY AS A METRIC?

#### 8.3.1.1 Analysis

Visualising the accuracy scores produced by detectors trained on balanced datasets of different sizes (Figure-10) the data suggests a close to linear relationship between the variables for all algorithms except Support Vector Machines, which seems to produce inconsistent results. This could be explained by the convergence issues highlighted above, and potentially makes the SVM results less valid.

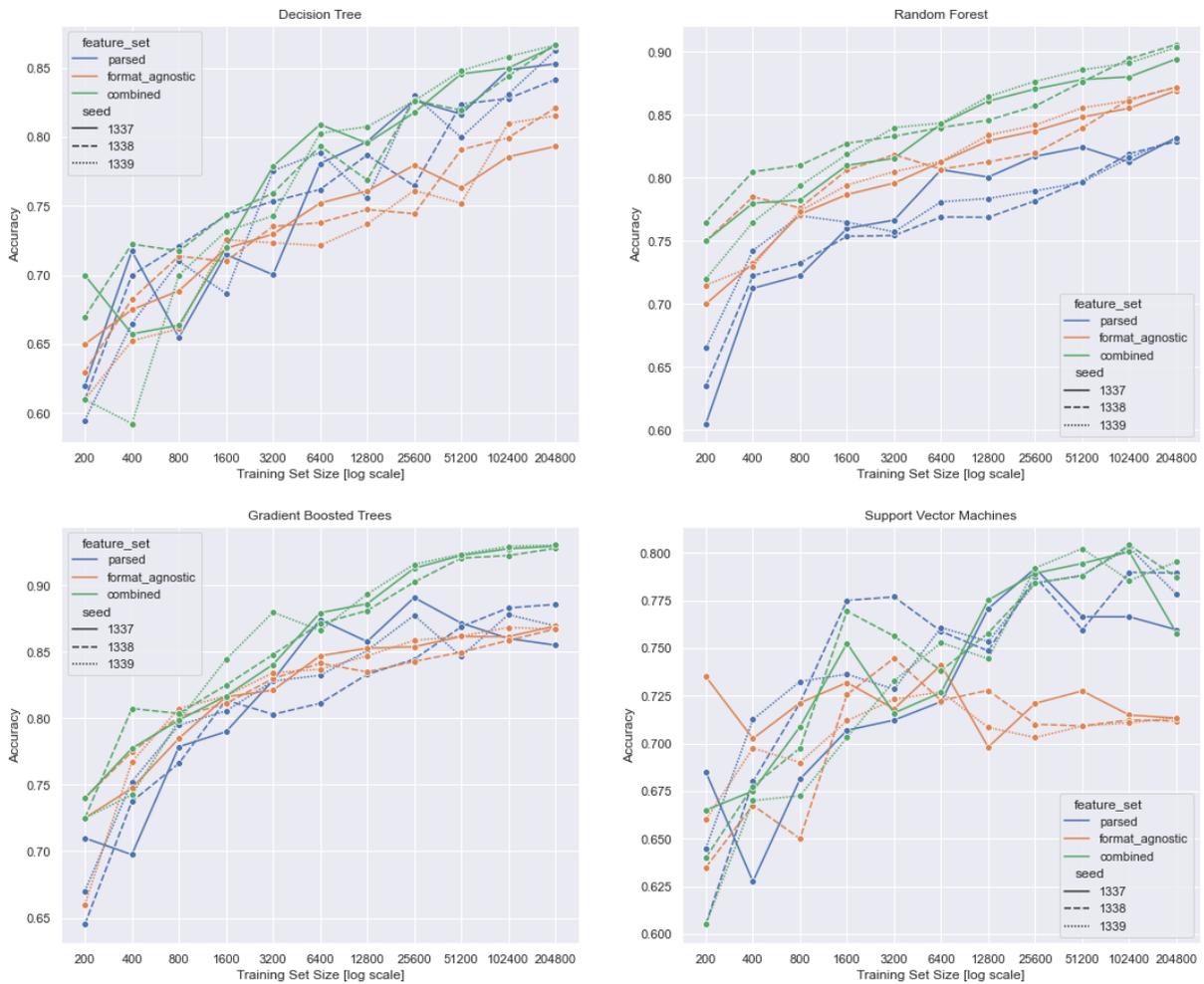

Figure-10: Accuracy scores for different training set sizes across detector types, feature sets and random seeds

Computing Pearson's correlation coefficients on the data (Table-7) the results show that there is a moderate correlation (Calkins, 2005) between accuracy and training set size for all tree-based models:

| Classifier Algorithm | Correlation Coefficient |
|---|---|
| Decision Tree | 0.646749 |
| Random Forest | 0.556330 |
| Gradient Boosted Trees | 0.515713 |
| Support Vector Machines | 0.381350 |

Table-7: Correlation coefficients between accuracy and training set size across classifier algorithms

Visualising the local Accuracy sensitivities (Figure-11) the data shows a moderate amount of noise with a suggestion of an underlying trend:

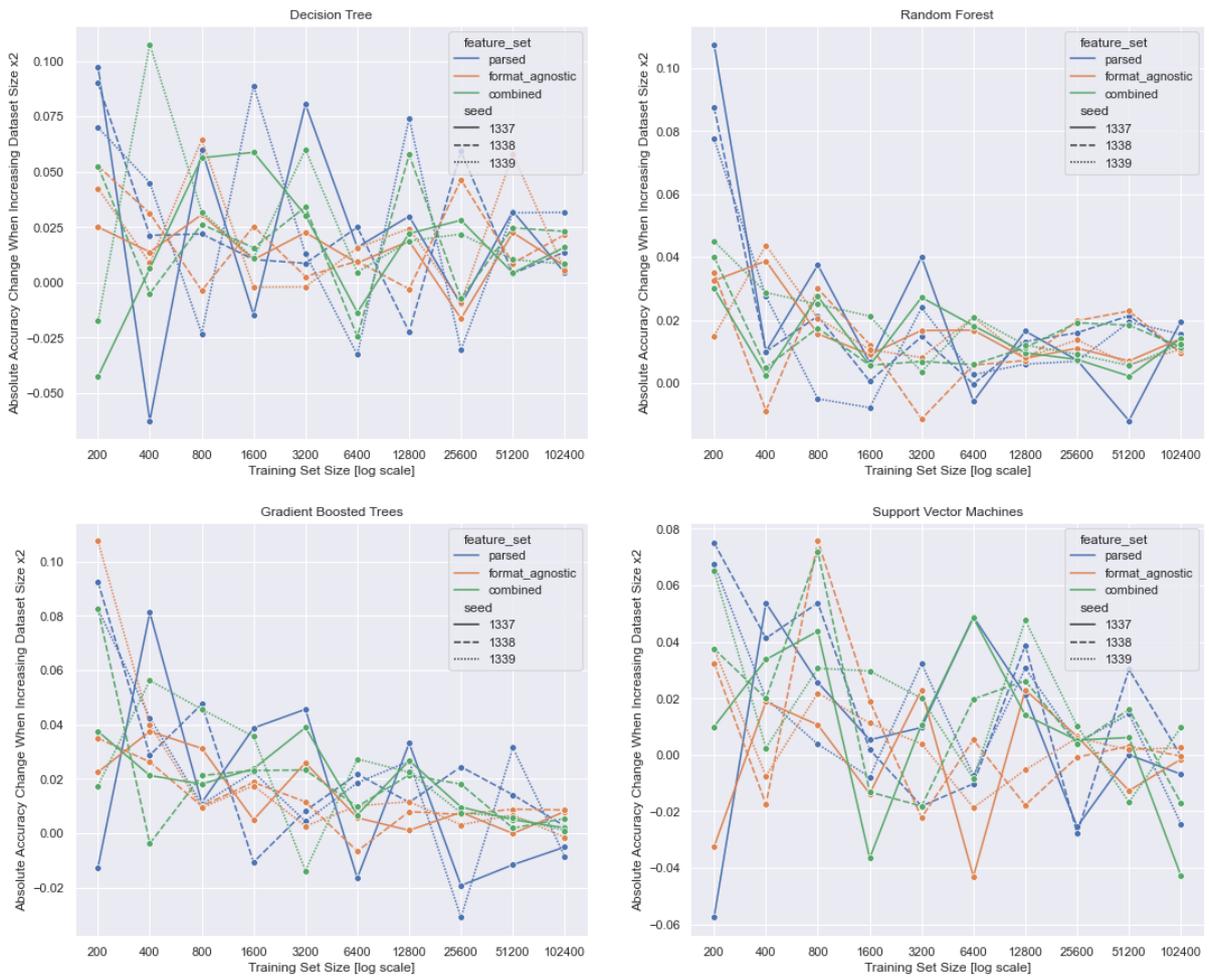

Figure-11: Accuracy sensitivities for different training set sizes across detector types, feature sets and random seeds

To better understand these trends, a regression plot could be used (Figure-12) that fits and displays a linear regression model on the data points:

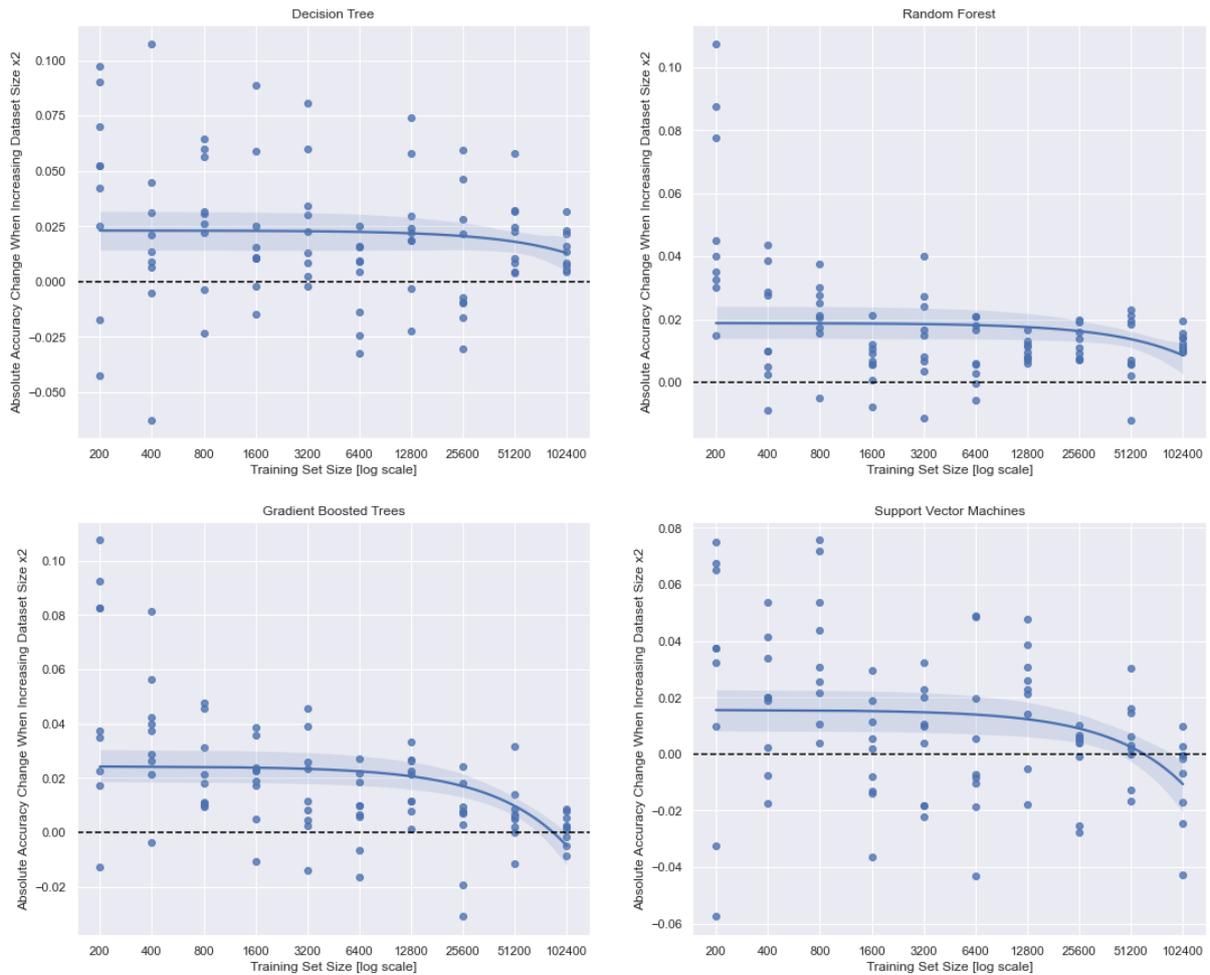
Figure-12: Accuracy sensitivity trends for different training set sizes across detector types, feature sets and random seeds

This regression suggests that there is a linear relationship between training set size and accuracy that results in a ~2% accuracy improvement each time the training set size is doubled until 12k-25k samples are reached, from which point the strength of the relationship starts to approach (and dips below) 0.

#### 8.3.1.2 Findings

The analysis has shown that all tree-based models produced detectors with 80%+ accuracy scores at 6400 samples (performance comparable to results in published literature), at which size the sensitivity was still high enough that increasing the dataset size yielded notable performance improvements. Using a large enough training set of 204 800 samples, the researcher successfully achieved a 93% accuracy score with a simple Gradient Boosted Tree model trained on the combined EMBER-2.0 feature set with default parameters.

Considering that in the literature differences between various author's results can be low single digit percentages, the data suggests that dataset size could be a factor significant enough to influence which technique produces better results and should be considered when performing comparisons.

## *8.3.2 QUESTION 2: WHAT IS THE SENSITIVITY OF DETECTOR PERFORMANCE TO DATASET SIZE WHEN TRAINED ON A BALANCED DATASET AND EVALUATED SIMULATING REAL-WORLD USAGE?*

### 8.3.2.1 Analysis

Figure-13 shows a visualisation of the "real-world" performance produced by detectors trained on balanced datasets of different sizes. Box plots were also provided to visualise the corresponding false positive rates, as the desired 1% FPR was not produced by Decision Tree and Support Vector Machines classifiers. Like for accuracy scores, the data suggests an almost linear relationship between training set size and performance except for the extremely noisy data produced by SVM measurements with known convergence issues:

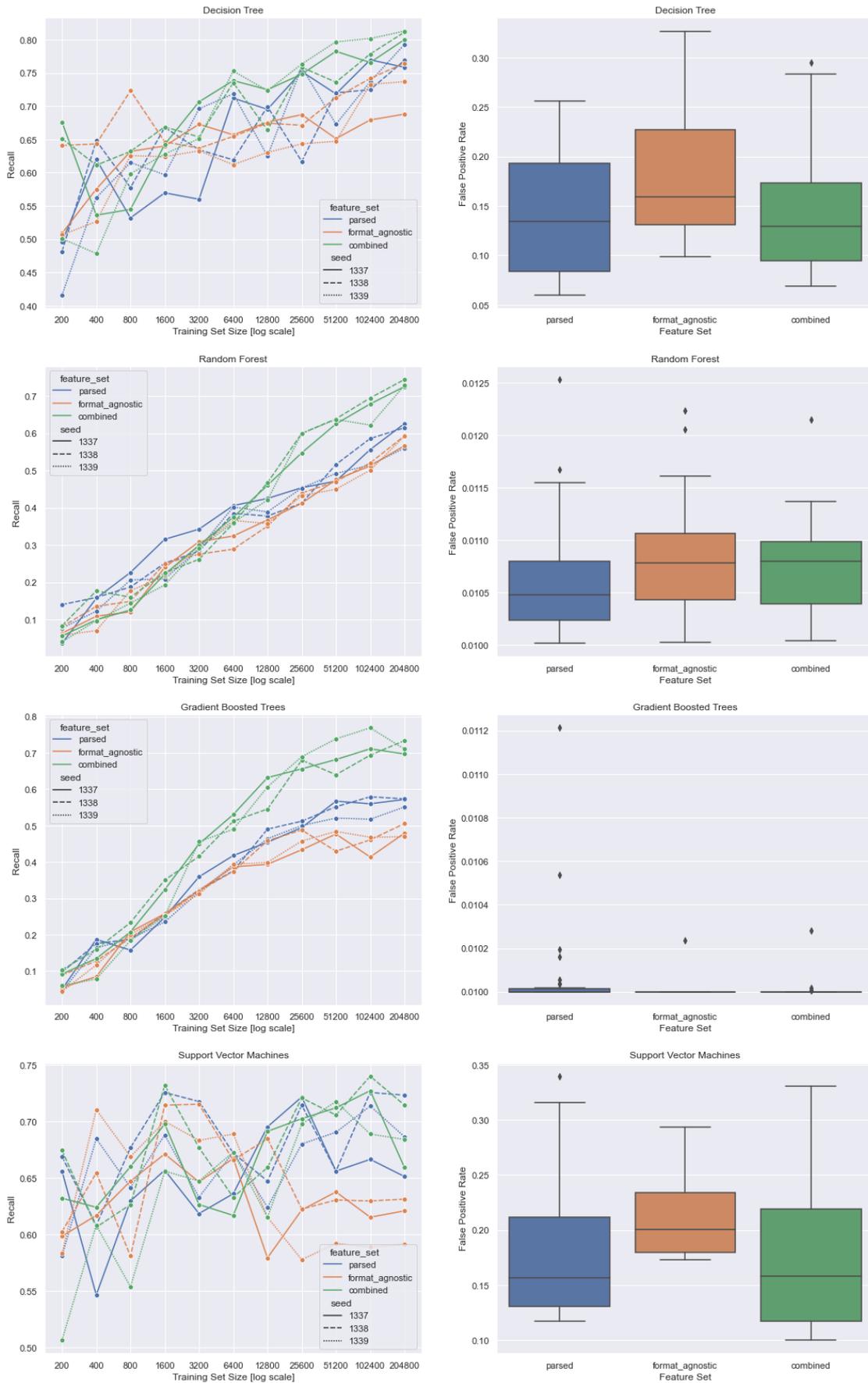

Figure-13: "Real-world" performance for different training set sizes across detector types, feature sets and random seeds and corresponding false positive rates.

Computing Pearson's correlation coefficients on the data (Table-8) reinforces this intuition for all tree-based models, showing moderate correlation between performance and training set size for Decision Trees and Gradient Boosted Trees, and high correlation for Random Forests:

| Classifier Algorithm | Correlation Coefficient |
| --- | --- |
| Decision Tree | 0.598539 |
| Random Forest | 0.740131 |
| Gradient Boosted Trees | 0.579187 |
| Support Vector Machines | 0.145247 |

Table-8: Correlation coefficients between real-world performance and training set size across classifier algorithms

Visualising the local Recall sensitivities (Figure-14) the data displays a significant amount of noise, however there is no obvious downward/upward trend apart from Gradient Boosted Trees, which data suggest a slight downward trend:

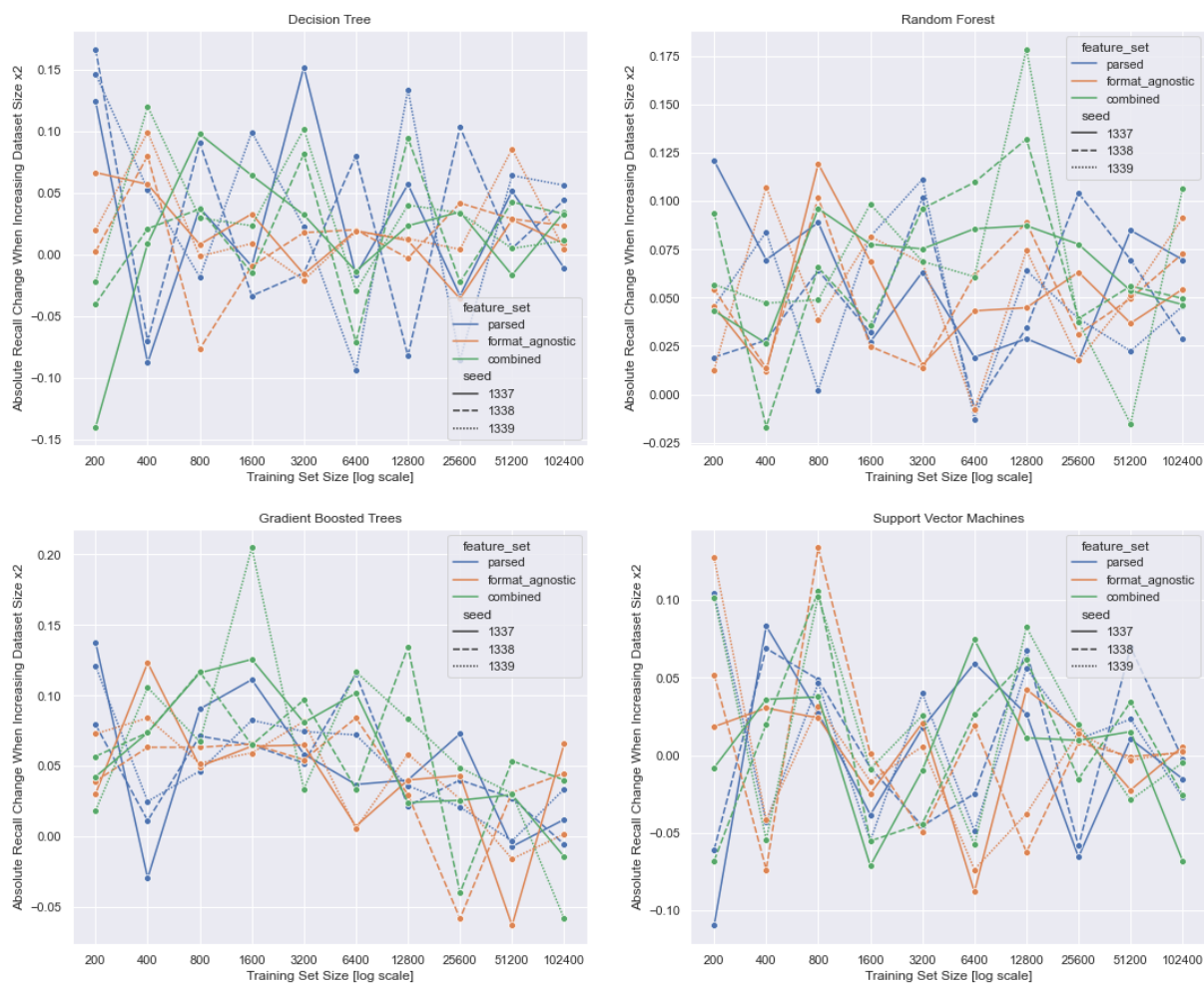

Figure-14: Recall sensitivities for different training set sizes across detector types, feature sets and random seeds

These observations are supported by applying a regression plot to the data points (Figure-15), which suggests that there is a strong linear relationship between training set size and recall (real-world performance) at a predetermined false positive rate for Decision Trees and Random

Forests contributing ~0.02 and ~0.05 improvements each time the training set is doubled. A similarly strong linear relationship is observed for Gradient Boosted Trees providing ~0.06 improvements until 12k samples are reached, from which point the strength of the relationship starts to approach 0.

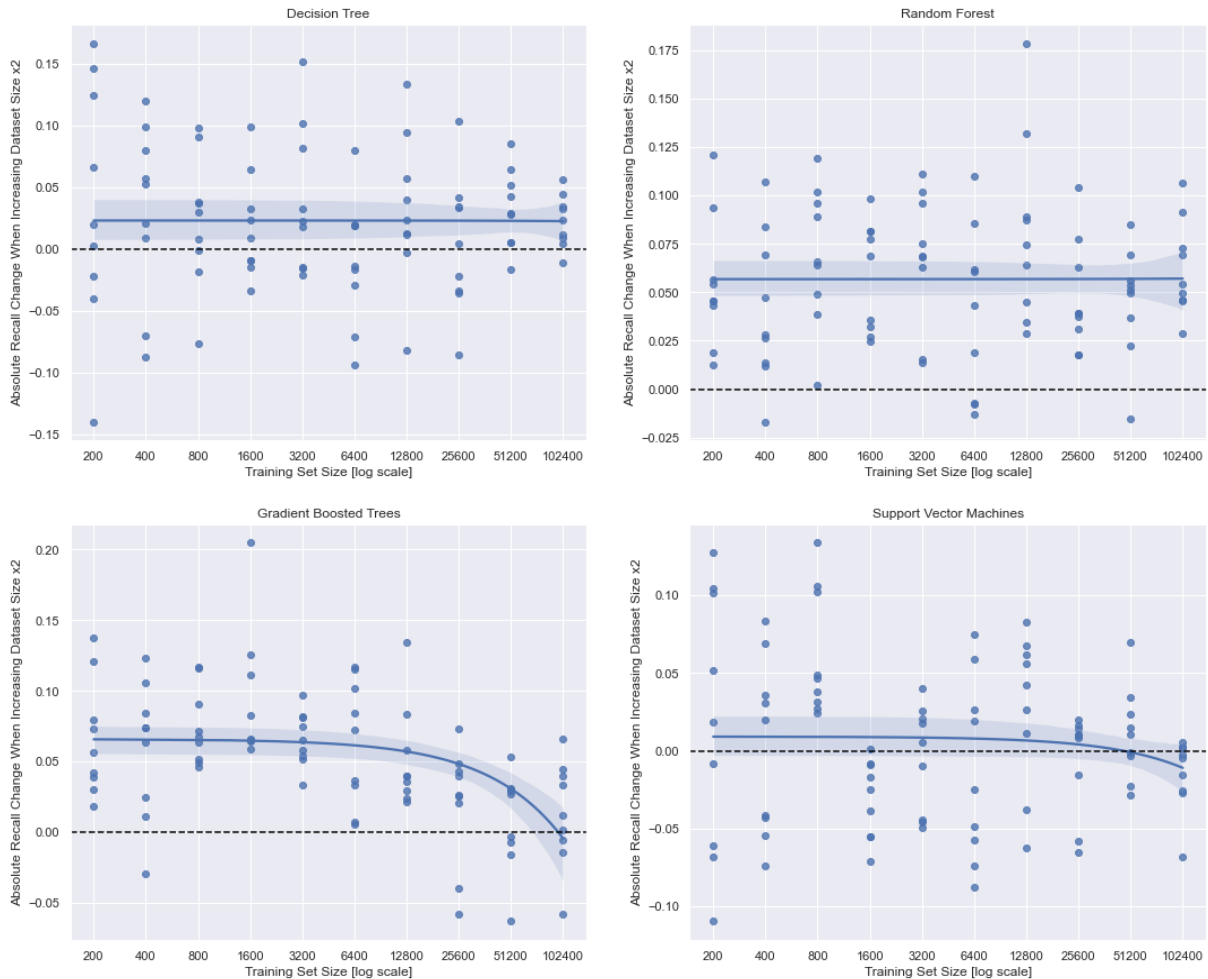

Figure-15: Recall sensitivity trends for different training set sizes across detector types, feature sets and random seeds

### 8.3.2.2 Findings

The analysis has shown that similarly to accuracy, "real-life" performance has a significant enough sensitivity to data set size for data set size to be a major contributor to performance when all other aspects of the experiment are fixed. The result support the finding from the question 1 results, that training set size should be a factor to consider when performing comparisons of published results.

The results also highlight the importance of chosen metrics. For example, Gradient Boosted Trees with 6400 samples on the feature agnostic dataset yielded accuracy scores approaching 85%, an impressive result, however, evaluating the same detector using the "real-life" performance measure the detector produced recall scores less than 0.4, indicating that the detector would catch less than 40% of all malwares in a realistic setting.

### *8.3.3 QUESTION 3: WHAT IS THE SENSITIVITY OF DETECTOR PERFORMANCE TO TESTING SET CLASS IMBALANCE WHEN TRAINED ON A BALANCED DATASET AND EVALUATED SIMULATING REAL-WORLD USAGE?*

#### 8.3.3.1 Analysis

Figure-16 shows a visualisation of the "real-world" performance produced by detectors trained on a balanced dataset of 204 800 samples and evaluated on testing datasets of varied malware/benignware ratios. Due to the experiment's varied false positive rate results, box plots were also provided to visualise corresponding false positive rates for context.

#### 8.3.3.2 Findings

This data shows a rather counter-intuitive but clear message. Some models seem to suffer a significant drop in performance as the number of benignware starts to increase from a 1:1 ratio, but for all models once the malware/benignware ratio exceeds 1:8, there is no significant change in "real-world" performance as the number of benignware is increased in the testing set. The data also shows, that for all tree-based models the false positive rates are also quite stable across the measurement points, however SVMs shows a greater variety, suggesting that even though the performance might be constant, false positive rates might vary. As SVMs had produced other notable problems during earlier steps this was not investigated further.

These results suggest that given a sufficiently large training set, depending on the modelling choices made, even balanced (1:1) testing sets could be suitable for evaluation, however once the imbalance exceeds 1:8 researchers should be confident that their performance metrics are representative of reality.

This is noteworthy, as given 5000 malware samples a 1:8 ratio only requires 40 000 benignware samples to perform testing, whereas simulating the 3% in-the-wild malware ratio as identified by Sophos (2021) would require > 150 000 benignware samples. Considering the difficulties associated with obtaining benignware (e.g.: copyright issues) this could be a significant reduction in the complexity of conducting research.

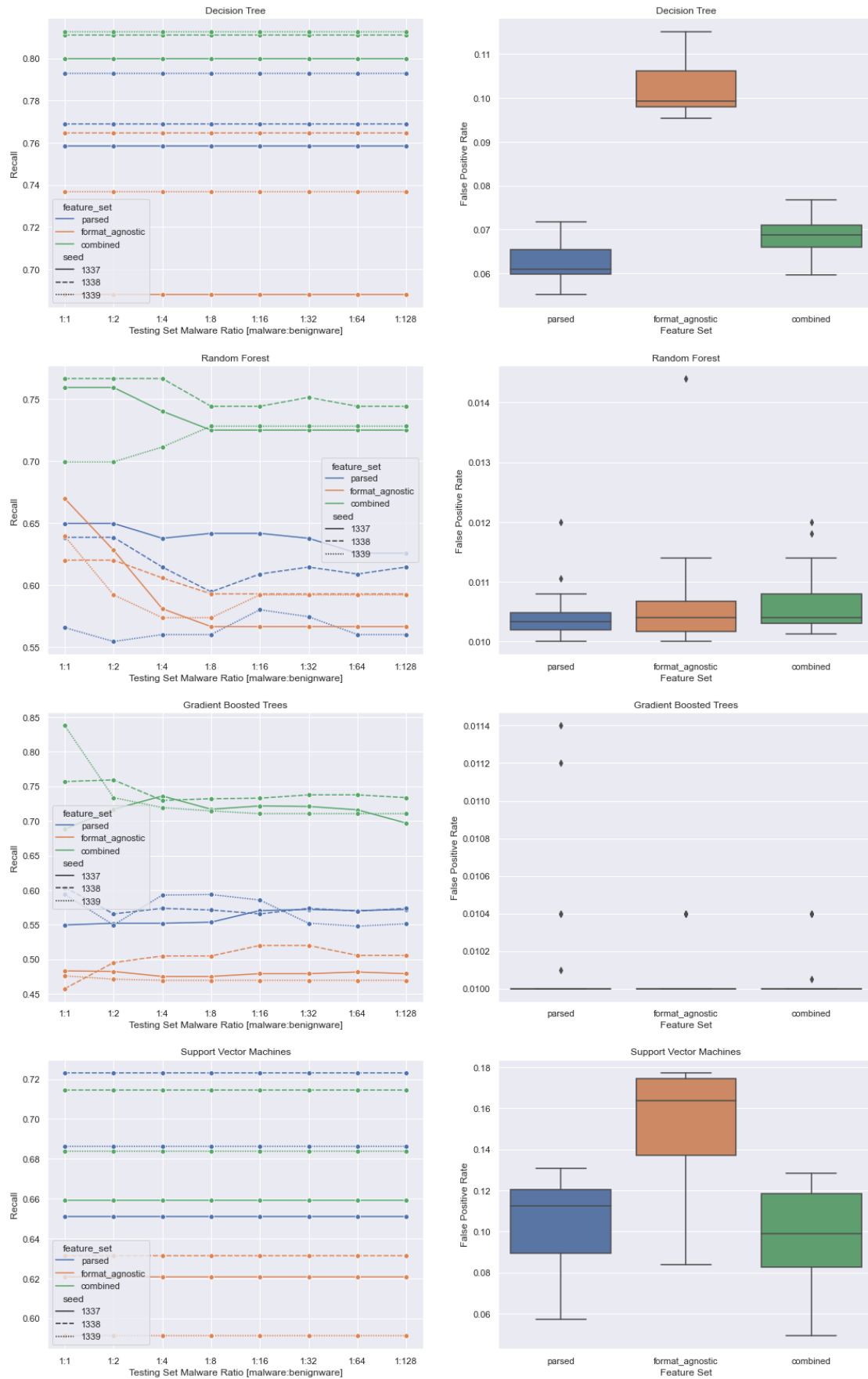

Figure-16: "Real-world" performance for various testing set malware ratios across detector types, feature sets and random seeds and corresponding false positive rates.

# 9 CONCLUSIONS

The research's aim was to investigate how results from published research on Windows machine learning-based detection techniques might be better analysed focusing on training dataset size and testing class imbalance regarding comparability and real-life applicability. The research has shown consistent results across tree-based models (Decision Trees, Random Forests, Gradient Boosted Trees) that both dataset size and testing class imbalance are factors significant enough in themselves to influence the comparability of results, and the trustworthiness of real-life applicability. Unfortunately, the results were restricted to tree-based models as Support Vector Machines data was deemed unreliable due to computation issues during experimentation.

Considering these computational problems, the researcher is of the opinion that the project achieved a moderate success overall, and a solid success in the reduced scope of tree-based models at fulfilling its aim.

The research's first objective was to understand if dataset size correlates to measured detector performance to an extent that prevents meaningful comparison of published results. The results from research question #1 (*What is the sensitivity of detector performance to dataset size when trained and evaluated on balanced datasets using accuracy as a metric?*) suggests that this might be the case. For all tree-based models the data has shown moderate correlation between accuracy and training set size for all feature sets (0.64, 0.55, 0.51 Pearson's correlation coefficients for Decision Trees, Random Forests, Gradient Boosted Trees respectively) with a closely linear relationship until 12k-25k samples are reached. Considering that papers surveyed by Singh and Singh (2021) reported various accuracy scores ranging from 83.42% to 99%, and that this research has successfully produced multiple detectors capable of increasing their accuracy from the ~80% accuracy range to 90+% accuracy solely by manipulating training set size, the research suggests that without understanding the nature of the training set size-accuracy curve for published results (e.g.: at which point performance plateau) conclusions between approaches on which approach is "better" shouldn't be made solely based on accuracy scores.

These results are consistent with Prusa *et al.*'s (2015) conclusions from the Tweet Sentiment problem domain, who found that increasing dataset size improves performance until a certain size, after which the performance gains diminish, and highlights the cross-domain importance of this phenomena.

The research's second objective was to understand if good performance reported in published research can be expected to perform well in a real-world deployment scenario. The performance of a "real-world" deployment scenario was interpreted as the recall metric (the fraction of examples classified as positive among the total number of positive) given a pre-determined false positive rate. The imbalance of malware/benignware in a real setting was also considered. The results suggest that high accuracy scores don't necessarily translate to high real-world performance. For example, evaluated Gradient Boosted Tree models trained on a format agnostic feature set of 204 800 samples, exceeding 85% accuracy across all random seeds, and couldn't even achieve a 0.5 recall depending on the random seed when evaluated at a 1% false positive rate regardless of the malware/benignware ratio in the test set. The research also produced counter examples, like Decision Trees trained on the combined feature set, achieving similar (~85%) accuracy scores, producing an impressive (>0.8) recall at around

~7% false positive rate, highlighting the importance of the usage of performance metrics more closely modelling real-world usage.

Results from question #2 (*What is the sensitivity of detector performance to dataset size when trained on a balanced dataset and evaluated simulating real-world usage?*) has shown similar/slightly higher correlation between "real-world" performance and training set size for all tree-based models (0.59, 0.74, 0.57 for Decision Trees, Random Forests, Gradient Boosted Trees respectively). Considering the accuracy-related results, this suggests that simply switching metrics from accuracy to recall at a fixed false positive rate, would not be sufficient on its own to facilitate a meaningful comparison.

Also, results from question #3 (*What is the sensitivity of detector performance to testing set class imbalance when trained on a balanced dataset and evaluated simulating real-world usage?*) has shown that as class imbalance changed from a 1:1 malware/benignware ratio when more benign test samples were added, measured "real-world" performance in some cases drastically changed until the ratio approaches 1:8, after which point the performance stabilised. This suggests that even if a more appropriate metric is used in combination with a high number of samples, the class imbalance ratio must be at least 1:8 to gain confidence that measured results will translate to similar performance during deployment.

These findings augment results from Roy *et al.* (2015) who performed similar research for Android malware and found that True Positive Rates (same as recall) and False Positive Rates were (more or less) constant as they increased malware/benignware ratios from 1:1 to 1:100 (through 1:5, 1:10, 1:20, 1:50) and highlight the importance of triangulation. Roy *et al.* only observed a single detector which is consistent with the fact that the performance degradation between 1:1 and 1:8 ratios was not observed in all cases by this research. It is recommended that similarly triangulated research should be performed for Android malware as well to confirm if similar recall degradation could be re-produced.

The research's findings are also consistent with results from Allix *et al.* (2016) who presented an approach for Android malware detection that exhibited high "in the lab" performance with low "in the wild" performance and argued that this performance gap might exist for other published approaches. This research has contributed multiple examples from the Windows malware detection problem domain that has shown similarly high "in the lab" measured accuracies with corresponding low real-world utility, strengthening Allix *et al.*'s argument that published approaches should be tested "in the wild" before their "significant" contribution to the malware detection domain could be accepted.

The above has implications both for academia, and cyber security practice more generally.

Researchers should be mindful of the impact of training set size and testing set imbalance on their research's comparability and applicability. The researcher recommends using sufficiently large training sets (at least 200 000 samples) and sufficiently imbalanced testing sets (at least 1:8 malware/benignware ratio) for future research, and/or the inclusion of multiple measurement points to understand the nature of the size/imbalance-to-performance curve. The researcher also recommends utilizing performance metrics that more closely mimic real-world deployment scenarios instead of accuracy. Unfortunately, this research could only provide evidence that existing practices could be problematic and could not give assurances on techniques that are guaranteed to be free from these problems. Also, the scope of the research was limited to features derived from static analysis and tree-based models. Because of these limitations, further similar research is recommended on other classifier (e.g.: Artificial Neural Networks) and feature types (e.g.: dynamic and hybrid approaches), and on potential practices that are free from the identified problems or have well understood risks. Further research should include investigation into identifying performance metrics that are the most suitable to mimic real-world deployment scenarios.

Industry practitioners when building home-grown detectors should primarily use the performance metric and class imbalance most representative of their deployment scenario when testing their approaches to maximize the relevance of their measurements, but also generate secondary results using metrics popular in research (e.g.: accuracy) on balanced testing sets to compare their results to published research. This (at the incremental cost of some added complexity) should help them bridge the identified potential gap between real-life applications and performance measured by contemporary research.

# APPENDIX

## APPENDIX I) SAMPLE OF THE GENERATED DATA

Full dataset available at: https://github.com/davidilles/msc-project-public/blob/main/results.csv

| | question | algorithm | feature_set | train_set_size | test_set_size | test_set_ratio | perf_measure | performance | other_info | seed |
|---|---|---|---|---|---|---|---|---|---|---|
| 951 | 3 | RF | combined | 204800 | 21250 | 1:16 | real-life | 0.724800 | 0.011050 | 1337 |
| 291 | 1 | DT | format_agnostic | 51200 | 51200 | 1:1 | accuracy | 0.763242 | NaN | 1337 |
| 217 | 1 | DT | parsed | 12800 | 12800 | 1:1 | accuracy | 0.786953 | NaN | 1338 |
| 754 | 2 | SVM | combined | 102400 | 161250 | 1:128 | real-life | 0.740000 | 0.108794 | 1338 |
| 784 | 2 | SVM | parsed | 204800 | 161250 | 1:128 | real-life | 0.723200 | 0.124887 | 1338 |
| 376 | 1 | RF | combined | 204800 | 204800 | 1:1 | accuracy | 0.906060 | NaN | 1338 |
| 268 | 1 | RF | combined | 25600 | 25600 | 1:1 | accuracy | 0.857148 | NaN | 1338 |
| 279 | 1 | SVM | parsed | 25600 | 25600 | 1:1 | accuracy | 0.791836 | NaN | 1337 |
| 667 | 2 | LGBM | parsed | 25600 | 161250 | 1:128 | real-life | 0.512000 | 0.010537 | 1338 |
| 122 | 1 | RF | format_agnostic | 1600 | 1600 | 1:1 | accuracy | 0.794375 | NaN | 1339 |
| 29 | 1 | SVM | parsed | 200 | 200 | 1:1 | accuracy | 0.645000 | NaN | 1339 |
| 453 | 2 | LGBM | format_agnostic | 400 | 161250 | 1:128 | real-life | 0.084800 | 0.010000 | 1337 |
| 277 | 1 | LGBM | combined | 25600 | 25600 | 1:1 | accuracy | 0.902656 | NaN | 1338 |
| 1056 | 3 | RF | format_agnostic | 204800 | 161250 | 1:128 | real-life | 0.566400 | 0.010469 | 1337 |
| 351 | 1 | SVM | parsed | 102400 | 102400 | 1:1 | accuracy | 0.766406 | NaN | 1337 |
| 960 | 3 | LGBM | combined | 204800 | 21250 | 1:16 | real-life | 0.721600 | 0.010000 | 1337 |
| 350 | 1 | LGBM | combined | 102400 | 102400 | 1:1 | accuracy | 0.929531 | NaN | 1339 |
| 813 | 3 | LGBM | format_agnostic | 204800 | 2500 | 1:1 | real-life | 0.483200 | 0.010400 | 1337 |
| 189 | 1 | RF | parsed | 6400 | 6400 | 1:1 | accuracy | 0.806562 | NaN | 1337 |
| 448 | 2 | RF | combined | 400 | 161250 | 1:128 | real-life | 0.176800 | 0.010063 | 1338 |
| 878 | 3 | RF | format_agnostic | 204800 | 6250 | 1:4 | real-life | 0.573600 | 0.011200 | 1339 |
| 195 | 1 | RF | combined | 6400 | 6400 | 1:1 | accuracy | 0.842812 | NaN | 1337 |
| 869 | 3 | DT | format_agnostic | 204800 | 6250 | 1:4 | real-life | 0.736800 | 0.098200 | 1339 |
| 509 | 2 | DT | format_agnostic | 1600 | 161250 | 1:128 | real-life | 0.624000 | 0.180062 | 1339 |
| 113 | 1 | DT | format_agnostic | 1600 | 1600 | 1:1 | accuracy | 0.725625 | NaN | 1339 |

## APPENDIX II) SOURCE CODE

Full codebase available at: https://github.com/davidilles/msc-project-public

### EMBER-2.0 FEATURE VECTORIZATION CODE

```
import os
import numpy as np
import pandas as pd
import json
from io import StringIO
import time
import hashlib

def current_milli_time():
    return round(time.time() * 1000)

pd.set_option('display.max_columns', None)
```

```python
data_dir = "/home/idomino/OU/t847/data/ember2018/"
out_dir = "/home/idomino/OU/t847/data/processed/"

def hash_string(s, m):
    return abs(hash(s)) % m

# Values
hist_size = 256
printdist_size = 96
hash_trick_dll = 128
hash_trick_import = 256
hash_trick_export = 128
hash_trick_sections = 50

coff_machines = ['AMD64', 'ARM', 'ARMNT', 'I386', 'IA64', 'MIPS16', 'MIPSFPU', 'POWERPC', 'R4000', 'SH3', 'SH4',
'THUMB']

coff_characteristics = ['AGGRESSIVE WS TRIM', 'BYTES REVERSED HI', 'BYTES REVERSED LO', 'CHARA 32BIT MACHINE',
                        'DEBUG_STRIPPED', 'DLL', 'EXECUTABLE_IMAGE', 'LARGE_ADDRESS_AWARE', 'LINE_NUMS_STRIPPED',
                        'LOCAL_SYMS_STRIPPED','NET_RUN_FROM_SWAP', 'RELOCS_STRIPPED', 'REMOVABLE_RUN_FROM_SWAP',
                        'SYSTEM', 'UP SYSTEM ONLY']

subsystems = ['EFI_APPLICATION', 'EFI_BOOT_SERVICE_DRIVER', 'EFI_RUNTIME_DRIVER', 'NATIVE', 'POSIX_CUI', 'UNKNOWN',
              'WINDOWS_BOOT_APPLICATION', 'WINDOWS_CE_GUI', 'WINDOWS_CUI', 'WINDOWS_GUI', 'XBOX']

dll_characteristics = ['APPCONTAINER', 'DYNAMIC_BASE', 'FORCE_INTEGRITY', 'GUARD_CF', 'HIGH_ENTROPY_VA',
                       'NO_BIND', 'NO_ISOLATION', 'NO_SEH', 'NX_COMPAT', 'TERMINAL_SERVER_AWARE', 'WDM_DRIVER']

magics = ['PE32', 'PE32_PLUS']

section_props = ['ALIGN_1024BYTES', 'ALIGN_128BYTES', 'ALIGN_16BYTES', 'ALIGN_1BYTES',
                 'ALIGN_2048BYTES', 'ALIGN_256BYTES', 'ALIGN_2BYTES', 'ALIGN_32BYTES',
                 'ALIGN_4096BYTES', 'ALIGN_4BYTES', 'ALIGN_512BYTES', 'ALIGN_64BYTES',
                 'ALIGN_8192BYTES', 'ALIGN_8BYTES', 'CNT_CODE', 'CNT_INITIALIZED_DATA',
                 'CNT_UNINITIALIZED_DATA', 'GPREL', 'LNK_COMDAT', 'LNK_INFO', 'LNK_NRELOC_OVFL',
                 'LNK_OTHER', 'LNK_REMOVE', 'MEM_16BIT', 'MEM_DISCARDABLE', 'MEM_EXECUTE', 'MEM_LOCKED',
                 'MEM_NOT_CACHED', 'MEM_NOT_PAGED', 'MEM_PRELOAD', 'MEM_READ',
                 'MEM_SHARED', 'MEM_WRITE', 'TYPE_NO_PAD']

# Appeared
header = ''
header += 'appeared'

# Histograms
for i in range(0,hist_size):
    header += f',histogram_{i}'
for i in range(0,hist_size):
    header += f',byteentropy_{i}'

# Strings
header += ',strings_num'
header += ',strings_avlength'
for i in range(0,printdist_size):
    header += f',strings_printabledist_{i}'
header += ',strings_printables'
header += ',strings_entropy'
header += ',strings_paths'
header += ',strings_urls'
header += ',strings_registry'
header += ',strings_MZ'

# General
header += ',general_size'
header += ',general_vsize'
header += ',general_has_debug'
header += ',general_exports'
header += ',general_imports'
header += ',general_has_relocations'
header += ',general_has_resources'
header += ',general_has_signature'
header += ',general_has_tls'
header += ',general_symbols'

# Header
header += ',header_coff_timestamp'
for machine in coff_machines:
    header += f',header_coff_machine_{machine}'
for characteristic in coff_characteristics:
    header += f',header_coff_{characteristic}'
for subsys in subsystems:
    header += f',header_opt_subsystem_{subsys}'
for characteristic in dll_characteristics:
    header += f',header_opt_ddl_characteristic_{characteristic}'
for magic in magics:
    header += f',header_opt_{magic}'
header += ',header_opt_major_image_version'
header += ',header_opt_minor_image_version'
header += ',header_opt_major_linker_version'
header += ',header_opt_minor_linker_version'
header += ',header_opt_major_operating_system_version'
```

```python
header += ',header_opt_minor_operating_system_version'
header += ',header_opt_major_subsystem_version'
header += ',header_opt_minor_subsystem_version'
header += ',header_opt_sizeof_code'
header += ',header_opt_sizeof_headers'
header += ',header_opt_sizeof_heap_commit'

# Sections
for i in range(0,hash_trick_sections):
    header += f',sections_h{i}_size'
    header += f',sections_h{i}_entropy'
    header += f',sections_h{i}_vsize'

for prop in section_props:
    header += f',sections_ENTRY_{prop}'

# Imports
for i in range(0,hash_trick_dll):
    header += f',imports_dll_h{i}_imported'
for i in range(0,hash_trick_import):
    header += f',imports_fun_h{i}_imported'

# Exports
for i in range(0,hash_trick_export):
    header += f',exports_h{i}'

# Control
header += ',label'
header += ',avclass'

def line_to_row(line):
    # Appeared
    row = ''
    data = json.loads(line)
    row += data['appeared'] + ','

    # Histograms
    for i in range(0,hist_size):
        row += str(data['histogram'][i]) + ','
    for i in range(0,hist_size):
        row += str(data['byteentropy'][i]) + ','

    # Strings
    row += str(data['strings']['numstrings']) + ','
    row += str(data['strings']['avlength']) + ','
    for i in range(0,printdist_size):
        row += str(data['strings']['printabledist'][i]) + ','
    row += str(data['strings']['printables']) + ','
    row += str(data['strings']['entropy']) + ','
    row += str(data['strings']['paths']) + ','
    row += str(data['strings']['urls']) + ','
    row += str(data['strings']['registry']) + ','
    row += str(data['strings']['MZ']) + ','

    # General
    row += str(data['general']['size']) + ','
    row += str(data['general']['vsize']) + ','
    row += str(data['general']['has_debug']) + ','
    row += str(data['general']['exports']) + ','
    row += str(data['general']['imports']) + ','
    row += str(data['general']['has_relocations']) + ','
    row += str(data['general']['has_resources']) + ','
    row += str(data['general']['has_signature']) + ','
    row += str(data['general']['has_tls']) + ','
    row += str(data['general']['symbols']) + ','

    # Header
    row += str(data['header']['coff']['timestamp']) + ','
    for machine in coff_machines:
        row += ('1' if data['header']['coff']['machine'] == machine else '0') + ','
    for characteristic in coff_characteristics:
        row += ('1' if characteristic in data['header']['coff']['characteristics'] else '0') + ','
    for subsys in subsystems:
        row += ('1' if data['header']['optional']['subsystem'] == subsys else '0') + ','
    for characteristic in dll_characteristics:
        row += ('1' if characteristic in data['header']['optional']['dll_characteristics'] else '0') + ','
    for magic in magics:
        row += ('1' if magic == data['header']['optional']['magic'] else '0') + ','
    row += str(data['header']['optional']['major_image_version']) + ','
    row += str(data['header']['optional']['minor_image_version']) + ','
    row += str(data['header']['optional']['major_linker_version']) + ','
    row += str(data['header']['optional']['minor_linker_version']) + ','
    row += str(data['header']['optional']['major_operating_system_version']) + ','
    row += str(data['header']['optional']['minor_operating_system_version']) + ','
    row += str(data['header']['optional']['major_subsystem_version']) + ','
    row += str(data['header']['optional']['minor_subsystem_version']) + ','
    row += str(data['header']['optional']['sizeof_code']) + ','
    row += str(data['header']['optional']['sizeof_headers']) + ','
    row += str(data['header']['optional']['sizeof_heap_commit']) + ','

    # Sections
    entry_section = None
    section_dict = {}
```

```python
        for i in data['section']['sections']:
            if i['name'] == data['section']['entry']:
                entry_section = i
            section_dict[hash_string(i['name'],hash_trick_sections)] = i

        for i in range(0,hash_trick_sections):
            section_data = section_dict.get(i)
            if section_data:
                row += str(section_data['size']) + ','
                row += str(section_data['entropy']) + ','
                row += str(section_data['vsize']) + ','
            else:
                row += '0,'
                row += '0,'
                row += '0,'

        if entry_section:
            entry_props = entry_section['props']
            for prop in section_props:
                row += ('1' if prop in entry_props else '0') + ','
        else:
            for prop in section_props:
                row += '0,'

        # Imports
        for i in range(0,hash_trick_dll):
            hashed_dlls = [hash_string(x, hash_trick_dll) for x in data['imports']]
            row += ('1' if i in hashed_dlls else '0') + ','
        for i in range(0,hash_trick_import):
            imported = False
            for key in data['imports']:
                hashed_funcs = [hash_string(f'{key}:{x}', hash_trick_import) for x in data['imports'][key]]
                if i in hashed_funcs:
                    imported = True
            row += ('1' if imported else '0') + ','

        # Exports
        for i in range(0,hash_trick_export):
            hashed_exports = [hash_string(x, hash_trick_export) for x in data['exports']]
            row += ('1' if i in hashed_exports else '0') + ','

        # Labels
        row += str(data['label']) + ','
        row += str(data['avclass']) if data['avclass'] else '-'

        return row

test_data = header
with open(data_dir + 'train_features_0.jsonl', 'r') as f:
    for i in range(0,5):
        line = f.readline()
        test_data += '\n'
        test_data += line_to_row(line)

df = pd.read_csv(StringIO(test_data))
print(df.dtypes)
df.columns

datafiles = ['train_features_0.jsonl', 'train_features_1.jsonl',
'train_features_2.jsonl','train_features_3.jsonl',
'train_features_4.jsonl','train_features_5.jsonl','test_features.jsonl']

def save_buffer(buffer, fragment):
    df = pd.read_csv(StringIO(header + buffer))
    df.to_pickle(f'{out_dir}data{fragment}.pkl', compression='zip')
    return ('', fragment+1)

t0 = current_milli_time()
buffer = ''
fragment = 0
chunksize = 50000
i = 0
for datafile in datafiles:
    print('Datafile:', datafile)
    with open(data_dir + datafile, 'r') as infile:
        while True:
            line = infile.readline()
            if not line:
                break
            row = line_to_row(line)
            buffer += '\n'
            buffer += row
            i += 1
            if i % chunksize == 0:
                t1 = current_milli_time()
                print(f'[{int((t1-t0)/1000)}]','Iteration:', i)
                buffer, fragment = save_buffer(buffer, fragment)
if buffer:
    save_buffer(buffer, fragment)
```

## SAMPLING CODE

```
import os
import numpy as np
import pandas as pd

pd.set_option('display.max_columns', 100)
pd.set_option('display.max_rows', 100)

input_dir = "/home/idomino/OU/t847/data/processed/"
output_dir = "/home/idomino/OU/t847/data/samples_new/"

seed_list = [1337, 1338, 1339]

for i in range(0,20):
    print(f'Reading dataframe #{i}...')
    df = pd.read_pickle(input_dir + f'data{i}.pkl', compression='zip')

    meta_df = df[['appeared','label','avclass']].copy()
    meta_df['index'] = meta_df.index
    meta_df['fragment'] = i

    mode = 'w' if i==0 else 'a'
    header = True if i==0 else False
    meta_df.to_csv(output_dir + 'metadata.csv', index=False, mode=mode, header=header)
    del df

metadata = pd.read_csv(output_dir + 'metadata.csv')
metadata.appeared = pd.to_datetime(metadata.appeared)

malware_mask = np.logical_and(metadata.avclass != '-',metadata.label == 1)
benign_mask = (metadata.label == 0)

first_malware_time = pd.Timestamp('2018-01-01 00:00:00')
split_time = pd.Timestamp('2018-07-31 00:00:00')

train_mask = np.logical_and(metadata.appeared >= first_malware_time, metadata.appeared < split_time)
test_mask = metadata.appeared > split_time

[np.logical_and(malware_mask,train_mask)].avclass.value_counts()[0:10]

metadata[np.logical_and(malware_mask,test_mask)].avclass.value_counts()[0:10]

top_n = 50
train_families = set(metadata[np.logical_and(malware_mask,train_mask)].avclass.value_counts()[0:top_n].index)
test_families = set(metadata[np.logical_and(malware_mask,test_mask)].avclass.value_counts()[0:top_n].index)
intersect_families = train_families.intersection(test_families)
print('Intersection families:', intersect_families)
print()
intersect_mask = metadata.avclass.apply(lambda x: x in intersect_families)

train_malware_samples = metadata[np.logical_and(malware_mask,np.logical_and(train_mask,intersect_mask))]
print('Train malware samples:', train_malware_samples.shape[0])

test_malware_samples = metadata[np.logical_and(malware_mask,np.logical_and(test_mask,intersect_mask))]
print('Test malware samples:', test_malware_samples.shape[0])

train_benign_samples = metadata[np.logical_and(benign_mask,train_mask)]
print('Train benign samples:', train_benign_samples.shape[0])

test_benign_samples = metadata[np.logical_and(benign_mask,test_mask)]
print('Test benign samples:', test_benign_samples.shape[0])

def prepare_samples(samples_for, n_malware, ratio_benign, seed):

    n_benign = n_malware * ratio_benign
    msg = f'[s{seed}] Preapring {samples_for} file of {n_malware} malware / {n_benign} benignware
(1:{ratio_benign})...'
    print(msg)

    malware_pool = None
    benign_pool = None

    if samples_for == 'train':
        malware_pool = train_malware_samples
        benign_pool = train_benign_samples
    elif samples_for == 'test':
        malware_pool = test_malware_samples
        benign_pool = test_benign_samples
    else:
        raise Exception('Invalid "sample_for" value, should be "train" or "test"!')

    malware_picked = malware_pool.sample(n_malware, random_state=seed)
    benign_picked = benign_pool.sample(n_benign, random_state=seed)

    acc_df = None
    for i in range(0,20):
        print(f'Reading dataframe #{i}...')
        df = pd.read_pickle(input_dir + f'data{i}.pkl', compression='zip')

        malware_idx = list(malware_picked[malware_picked.fragment == i]['index'])
        benign_idx = list(benign_picked[benign_picked.fragment == i]['index'])
```

```
        idx = malware_idx + benign_idx

        if acc_df is not None:
            acc_df = pd.concat([acc_df,df.loc[idx].copy()])
        else:
            acc_df = df.loc[idx].copy()

        del df

    core_columns = ['appeared', 'label', 'avclass']
    feature_columns = [x for x in acc_df.columns if x not in core_columns]
    format_agnostic_columns = [x for x in feature_columns
                                 if x.startswith('histogram')
                                 or x.startswith('byteentropy')
                                 or x.startswith('strings')]
    parsed_columns = [x for x in feature_columns if x not in format_agnostic_columns]

    format_agnostic_columns = format_agnostic_columns + core_columns
    parsed_columns = parsed_columns + core_columns

    format_agnostic_df = acc_df[format_agnostic_columns]
    parsed_df = acc_df[parsed_columns]

    filename = f'{samples_for}_{n_malware}_malware_x{ratio_benign}_benign_format_agnostic_s{seed}.pkl'
    format_agnostic_df.to_pickle(output_dir + filename, compression='zip')
    print('Saved:', filename)

    filename = f'{samples_for}_{n_malware}_malware_x{ratio_benign}_benign_parsed_s{seed}.pkl'
    parsed_df.to_pickle(output_dir + filename, compression='zip')
    print('Saved:', filename)

    filename = f'{samples_for}_{n_malware}_malware_x{ratio_benign}_benign_combined_s{seed}.pkl'
    acc_df.to_pickle(output_dir + filename, compression='zip')
    print('Saved:', filename)

required_samples = [
    ('train', 100, 1),
    ('train', 200, 1),
    ('train', 400, 1),
    ('train', 800, 1),
    ('train', 1600, 1),
    ('train', 3200, 1),
    ('train', 6400, 1),
    ('train', 12800, 1),
    ('train', 25600, 1),
    ('train', 51200, 1),
    ('train', 102400, 1),

    ('test', 100, 1),
    ('test', 200, 1),
    ('test', 400, 1),
    ('test', 800, 1),
    ('test', 1600, 1),
    ('test', 3200, 1),
    ('test', 6400, 1),
    ('test', 12800, 1),
    ('test', 25600, 1),
    ('test', 51200, 1),
    ('test', 102400, 1),

    ('test', 1250, 1),
    ('test', 1250, 2),
    ('test', 1250, 4),
    ('test', 1250, 8),
    ('test', 1250, 16),
    ('test', 1250, 32),
    ('test', 1250, 64),
    ('test', 1250, 128)
]

for sample_params in required_samples:
    for s in seed_list:
        prepare_samples(*sample_params, s)
```

## DATA GENERATION CODE

```
import numpy as np
import pandas as pd
from sklearn.pipeline import make_pipeline
from sklearn.preprocessing import StandardScaler
from sklearn.tree import DecisionTreeClassifier
from sklearn.ensemble import RandomForestClassifier
from sklearn.svm import SVC
from sklearn.svm import LinearSVC
import lightgbm as lgb
from sklearn.metrics import accuracy_score
from sklearn.metrics import recall_score
from sklearn.metrics import roc_auc_score
from sklearn.metrics import roc_curve
from sklearn.metrics import plot_roc_curve
```

```python
import matplotlib.pyplot as plt
import pickle
import datetime

data_dir = "/home/idomino/OU/t847/data/samples_new/"
model_dir = "/home/idomino/OU/t847/data/models_new/"
results_dir = "/home/idomino/OU/t847/data/results_new/"

def get_dataset(n_malware, ratio_benign, dataset_type, features, seed):

    filename = f'{dataset_type}_{n_malware}_malware_x{ratio_benign}_benign_{features}_s{seed}.pkl'

    df = pd.read_pickle(data_dir + filename, compression='zip')
    print('Read sample:', filename)

    feature_columns = [x for x in df.columns if x not in ['appeared','label','avclass']]

    X = df[feature_columns]
    y = df['label']

    return (X, y, df)

def get_classifier(clf, seed):
    if clf == 'DT':
        return DecisionTreeClassifier(random_state=seed)
    elif clf == 'RF':
        return RandomForestClassifier(random_state=seed)
    elif clf == 'SVM':
        return make_pipeline(StandardScaler(), LinearSVC(random_state=seed))
    elif clf == 'LGBM':
        return lgb.LGBMClassifier(random_state=seed)
    else:
        return None

def train_and_save_model(n_malware, ratio_benign, features, clf_type, seed):
    X, y, df = get_dataset(n_malware, ratio_benign, 'train', features, seed)
    print(f'[{datetime.datetime.now()}]', 'Loaded X', X.shape, 'and y', y.shape)
    clf = get_classifier(clf_type, seed)
    print(f'[{datetime.datetime.now()}]','Starting fit:')
    clf.fit(X,y)
    print(f'[{datetime.datetime.now()}]','Fitted',clf)

    filename = f'{clf_type}_{n_malware}_malware_x{ratio_benign}_benign_{features}_s{seed}.pkl'
    pickle.dump(clf, open( model_dir + filename, "wb" ))
    print(f'[{datetime.datetime.now()}]','Saved model:', filename)
    print()
    # Clean up
    del X
    del y
    del df

for n_malware in [100, 200, 400, 800, 1600, 3200, 6400, 12800, 25600, 51200, 102400]:
    for clf_type in ['DT', 'RF', 'LGBM','SVM']:
        for features in ['parsed', 'format_agnostic', 'combined']:
            for seed in [1337, 1338, 1339]:
                train_and_save_model(n_malware, 1, features, clf_type, seed)

def create_observation(question, train_n_malware, test_n_malware, test_ratio_benign,
                       features, clf_type, metric, seed):

    model_pkl = f'{clf_type}_{train_n_malware}_malware_x1_benign_{features}_s{seed}.pkl'
    model = pickle.load(open(model_dir + model_pkl,"rb"))
    print('Loaded model', model, 'from:', model_pkl)

    X, y, df = get_dataset(test_n_malware, test_ratio_benign, 'test', features, seed)
    y_pred = None
    y_score = None
    retval = {
        'question': question, 'algorithm': clf_type, 'feature_set': features,
        'train_set_size': train_n_malware * 2,
        'test_set_size': test_n_malware + test_n_malware * test_ratio_benign,
        'test set ratio': f'1:{test_ratio_benign}',
        'perf_measure': metric,
        'seed': seed
    }

    if metric == 'accuracy':
        y_pred = model.predict(X)
        acc = accuracy_score(y, y_pred)
        retval['performance'] = acc
        retval['other_info'] = None
    elif metric == 'AUC' and clf_type != 'SVM':
        y_score = model.predict_proba(X)[:,1]
        auc = roc_auc_score(y, y_score)
        retval['performance'] = auc
        retval['other_info'] = None
    elif metric == 'AUC' and clf_type == 'SVM':
        retval['performance'] = None
        retval['other_info'] = None
    elif metric == 'real-life' and clf_type != 'SVM':
        y_score = model.predict_proba(X)[:,1]
        fpr,tpr,thresholds = roc_curve(y, y_score, drop_intermediate=False)
        i = np.argmax(fpr>=0.01)
```

```python
            retval['performance'] = tpr[i]
            retval['other_info'] = fpr[i]
    elif metric == 'real-life' and clf_type == 'SVM':
        y_pred = model.predict(X)
        FP = np.logical_and(y == 0, y_pred == 1).sum()
        fpr = FP/len(y)
        tpr = recall_score(y, y_pred)
        retval['performance'] = tpr
        retval['other_info'] = fpr
    else:
        raise Exception('Invalid metric:', metric)

    # Clean up
    del X
    del y
    del df
    del model
    if y_pred is not None:
        del y_pred
    if y_score is not None:
        del y_score

    return retval

observations = []

# Question 1:
for n_malware in [100, 200, 400, 800, 1600, 3200, 6400, 12800, 25600, 51200, 102400]:
    for clf_type in ['DT', 'RF', 'LGBM','SVM']:
        for features in ['parsed', 'format_agnostic', 'combined']:
            for seed in [1337, 1338, 1339]:
                obs = create_observation(1, n_malware, n_malware, 1, features, clf_type, 'accuracy', seed)
                observations.append(obs)

# Question 2:
for n_malware in [100, 200, 400, 800, 1600, 3200, 6400, 12800, 25600, 51200, 102400]:
    for clf_type in ['DT', 'RF', 'LGBM','SVM']:
        for features in ['parsed', 'format_agnostic', 'combined']:
            for seed in [1337, 1338, 1339]:
                obs = create_observation(2, n_malware, 1250, 128, features, clf_type, 'real-life', seed)
                observations.append(obs)

# Question 3
for benign_ratio in [1, 2, 4, 8, 16, 32, 64, 128]:
    for clf_type in ['DT', 'RF', 'LGBM', 'SVM']:
        for features in ['parsed', 'format_agnostic', 'combined']:
            for seed in [1337, 1338, 1339]:
                obs = create_observation(3, 102400, 1250, benign_ratio, features, clf_type, 'real-life', seed)
                observations.append(obs)

# Create and save dataframe
observations_df = pd.DataFrame(observations, columns=['question', 'algorithm', 'feature_set',
                                    'train_set_size', 'test_set_size', 'test_set_ratio',
                                    'perf_measure', 'performance', 'other_info', 'seed'])

observations_df.to_csv(results_dir + 'results.csv', index=False)
```